\documentclass[useAMS,usenatbib]{mn2e}
\usepackage{amssymb,rotating,natbib,float,longtable,array}
\usepackage[T1]{fontenc}
\usepackage{aecompl}
\newcommand{\msun}{\ifmmode \,\mathrm{M_{\odot}} \else M$_{\odot}\quad$\fi}
\newcommand{\mhesun}{\ifmmode \,\mathrm{M_{\odot}} \else M$_{\odot,\mathrm{He}}\quad$\fi}
\newcommand{\rsun}{\ifmmode R_{\odot} \else R$_{\odot}$\fi}
\newcommand{\lsun}{\ifmmode L_{\odot} \else L$_{\odot}$\fi}
\newcommand{\zsun}{\ifmmode Z_{\odot} \else $Z_{\odot}$\fi}

\graphicspath{ {./images/} }
\DeclareGraphicsExtensions{.eps}
\title[Helium stars]
      {Helium Stars: Towards an understanding of Wolf--Rayet Evolution}
\author[L. A. S. McClelland and J. J. Eldridge]
{L. A. S. McClelland$^1$, J.J. Eldridge$^1$\\
\\
$^{1}$The Department of Physics, The University of Auckland, Private Bag 92019, Auckland, New Zealand}

\begin{document}

\maketitle		

\begin{abstract}

Wolf--Rayet (WR) stars are massive stars that have lost most or all of their hydrogen via powerful stellar winds. Recent observations have indicated that hydrogen-free WR stars have cooler temperatures than those predicted by current evolutionary models. To investigate how varying mass-loss rate affects WR evolution, we have created a grid of pure helium star models. We compare our results with Galactic and LMC WR observations and show that the temperature ranges of observed WR stars can be reproduced by varying the mass-loss rate, which effectively determines the size of the helium envelope around the core. We also find that WN and WO stars arise from more massive stars, whereas WC stars come from lower masses. This contradicts the standard Conti scenario by which WN and WC stars evolve in an age sequence. We also predict the magnitudes of our models at core-collapse and compare with observations of nearby progenitors of Type Ib/c supernovae. We confirm the findings of previous studies that suggest WR stars are the progenitors of core-collapse supernovae: the progenitors would remain unobserved except in the cases where the progenitor is a low-mass helium giant, as is the case of iPTF13bvn.
\end{abstract}


\section{Introduction}
\label{sec:intro}
Massive stars are the most charitable of their kind. Throughout their relatively short lives, they continually donate nucleosynthetic materials to the surrounding interstellar medium (ISM). Then, upon death, further give to the ISM with a final act of philanthropy: a violent explosion known as a core-collapse supernova (CCSN) \citep[][]{2003ApJ...591..288H,2012ARA&A..50..107L}. These supernovae (SNe) are grouped according to their light curves and spectra. Broadly, SNe with a presence of hydrogen are classed Type II; those absent hydrogen are classed Type Ib or Ic, depending on the presence or absence of helium, respectively \citep[][]{1997ARA&A..35..309F}\footnote{There exists another class of Type I SN, a thermonuclear SN labelled Type Ia. Though not considered further within this paper, they are discussed by \citet{2012NewAR..56..122W} and \citet{2013FrPhy...8..116H}}. 

Wolf--Rayet (WR) stars are, classically, massive helium-burning stars that, through strong mass loss, have lost all or most of their hydrogen envelopes leaving a partially or fully exposed helium core. WR stars are classified by the spectral presence of specific burning products on their surface: a nitrogen sequence, WN, with the presence of nitrogen; a carbon sequence, WC, with the presence of carbon; or an oxygen sequence, WO, with the presence of oxygen. The former two may be further subdivided into early- or late-types--WNE (WCE) or WNL (WCL), respectively--depending on the surface temperatures of the stars \citep[][]{2007ARA&A..45..177C}.

WR stars are characterised by broad emission lines, which indicate the presence of a dense, optically thick, high-velocity wind. A consequence of strong mass loss is that most WR stars will end their lives free of hydrogen. Hence, we may anticipate the observed supernova spectra of such stars will also be free of hydrogen. Thus, WR stars are favourable candidates for progenitors of Type Ib/c supernovae.

The general picture of WR evolution is known as the ``Conti scenario'' \citep{conti1975relationship}. Conti proposed a scenario for massive O stars whereby the stage of evolution can be spectroscopically sequenced from the revealed burning products.

With the Conti scenario, we may describe the evolution of a main-sequence star in terms of its initial mass \citep[see][]{2007ARA&A..45..177C}. For an initial mass in the range $\sim 25-40\msun$,
{\small
	\[ \mathrm{O} \rightarrow \mathrm{LBV/RSG} \rightarrow \mathrm{WN(H-poor)} \rightarrow \mathrm{SN\, Ib}; \]
}\noindent alternatively, for initial mass within $\sim 40-75\msun$,
{\small
	\[ \mathrm{O} \rightarrow \mathrm{LBV} \rightarrow \mathrm{WN(H-poor)} \rightarrow \mathrm{WC} \rightarrow \mathrm{SN\, Ic}; \]
}\noindent and for masses $\gtrsim 75\msun$
{\small
\[ \mathrm{O} \rightarrow \mathrm{WN(H-rich)} \rightarrow \mathrm{LBV} \rightarrow \mathrm{WN(H-poor)} \rightarrow \mathrm{WC} \rightarrow \mathrm{SN\, Ic}. \]
}

\begin{figure}
	\includegraphics[width=\columnwidth]{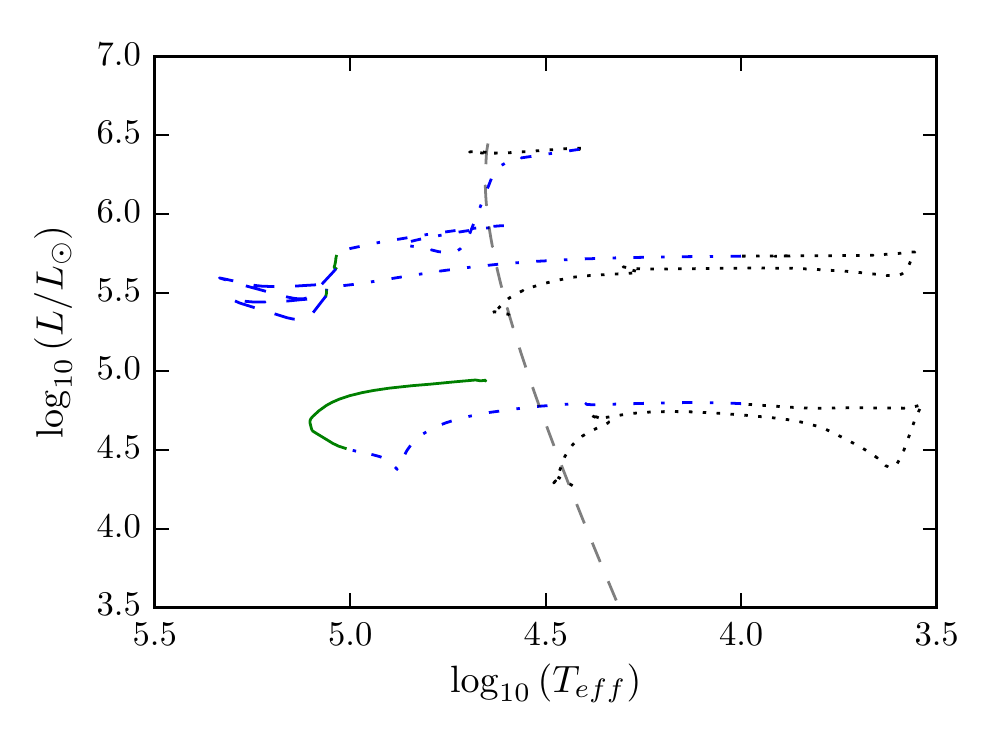}
	\caption[]{Evolutionary comparison of a $150\msun$ model, a $40\msun$ model, and a $15\msun$ binary model from the hydrogen zero-age main sequence (indicated). The line styling of the tracks denotes the type of mass loss applied: dotted, broken-dotted, solid and broken represent pre-WR, WNL, WN and WC, respectively.}
	\label{fig:HRComp}
\end{figure}

\begin{figure}
	\centering
	\includegraphics[width=\columnwidth]{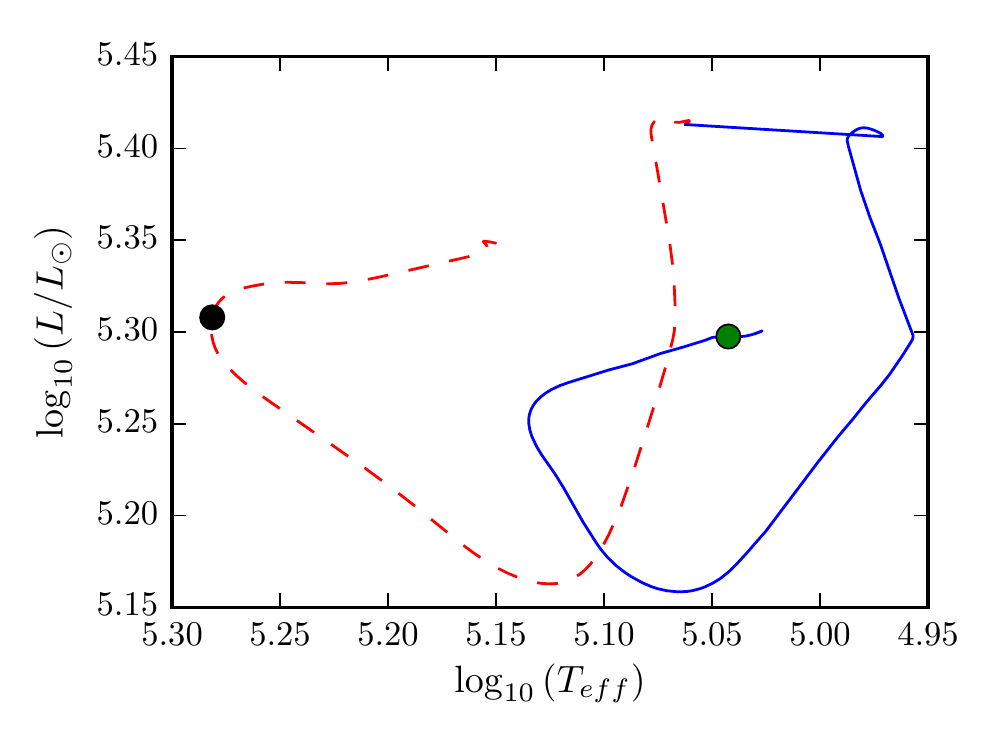}
	\caption[]{HR diagram showing the evolutionary tracks (see text) of a pure helium star with an initial mass of $14\msun$ at solar metallicity with (solid, blue) and without (broken, red) clumping. The ignition point of shell helium is marked on each track. Final stellar masses are $7.7\msun$ and $7.8\msun$ for unclumped and clumped, respectively.}
	\label{fig:WRBurningPhases}
\end{figure}

Fig.~\ref{fig:HRComp} shows the evolution on the Hertzsprung--Russell (HR) diagram of three massive hydrogen zero-age main-sequence (HZAMS) models from the ignition of core hydrogen to the start of core-neon burning. Both models enter the WR phase near the start of core-helium burning, the helium zero-age main-sequence (HeZAMS). From the HeZAMS, with a now hydrogen-free model, core-helium burning is initiated and the evolution proceeds as in Fig.~\ref{fig:WRBurningPhases}. We allow variation of the mass-loss rates to account for any potential helium consumption prior to the WR phase. We note the evolutionary similarity between Fig.~\ref{fig:HRComp} and the post-helium-ignition portion of Fig.~\ref{fig:WRBurningPhases}, allowing us to begin at the HeZAMS and avoiding any uncertainties from pre-WR evolution.

The detailed analysis of Galactic WN stars performed by \citet{2006A&A...457.1015H} showed the stellar temperatures of observed hydrogen-free WN stars were much cooler than temperatures predicted by evolutionary models.
The work of \citet{1999PASJ...51..417I, 2006A&A...450..219P} revealed massive helium stars possess an extended envelope above a compact core. The severity of this radial inflation is dependent on the metallicity by way of the iron-opacity peak \citep{2012A&A...538A..40G}. The effect is further enhanced with density inhomogeneities (``clumping'') in the extended envelope. As the outer envelope is extended (sometimes by orders of magnitude), the result is a reduction of the effective temperature. We show how clumping changes the evolution of a model in Fig.~\ref{fig:WRBurningPhases}. Therefore, a study into the evolution of these stars should be undertaken.  

In their study of Galactic WC stars, \citet{2012A&A...540A.144S} concluded that the initial mass range for WC stars is between 20 and 45 solar masses--significantly less than previously thought. With our grids of helium star models, we provide an evolutionary test of this conclusion.

Previous theoretical studies on the evolution of WR stars have considered the effects of rotation \citep{2012A&A...542A..29G,2005A&A...429..581M} and duplicity \citep{1998NewA....3..443V,2003A&A...400...63V,2007ApJ...662L.107V,2013MNRAS.436..774E}.
Following \citet{1986A&A...167...61H} and \citet{2002MNRAS.331.1027D}, we shall investigate the evolution of single helium star models. We extend and update their work by considering higher masses of helium stars and varying the input physics--initial metallicity and mass-loss rate--of our models.

In this paper, we shall investigate the evolution of pure helium stars. We begin with an overview of the calculation of the models and their properties. We then evolve the models and analyse the results at various stages of evolution; in particular, the structure and composition of the stars. Finally, we compare our findings with current observational data in an extensive discussion.

\section{Computational Method}
\label{sec:nummeth}

To investigate the evolution of helium stars, we have constructed a grid of hydrogen-free models.
We make use of the Cambridge \textsc{stars} evolutionary code. Originally developed by \citet{1971MNRAS.151..351E}, it has been modified by various groups; herein, we employ the version described by \citet{2009MNRAS.396.1699S}.

We create our grid of pure helium models by evolving a $30\msun$ model from the HZAMS with strong, constant mass loss whilst allowing only hydrogen consumption; evolution is halted once the model reaches approximately one solar mass. The model is then evolved with the burning of helium allowed, but the consumption of helium suppressed (effectively eliminating any chemical change by way of helium burning), so a stable helium model is resultant. We then successively add mass to or remove mass from the model to obtain a grid of helium models of various masses. The models are relaxed to chemical homogeneity and thermal equilibrium before evolution is initiated. 

\subsection{Metallicity}
\label{ssec:construction}

We make our selection of metallicities based on the expected environments of WR stars: $Z=0.008$ for the Large Magellanic Cloud; $Z=0.014$ and $Z=0.02$ being, respectively, ``new'' and ``old'' solar metallicity; and $Z=0.04$, double ``old'' solar. The situation regarding the exact value of solar metallicity is currently confused. We have used \citet{1998SSRv...85..161G} for the ``old'' solar composition. For the ``new'' solar composition, we have used \citet{2009ARA&A..47..481A}, which is consistent with \citet{2012A&A...539A.143N}. We are, however, constrained to using the opacity tables of \citet{2004MNRAS.348..201E}, which use the older solar compositions and are more oxygen rich than more modern solar compositions. Regardless, for Wolf-Rayet stars the iron-abundance is of primary importance. To evaluate the accuracy of this choice, we list the compositions of our models in Table~\ref{tab:ohfe}.

We note here a caveat: our models do not have hydrogen in them, so slightly different [O/H] values can be obtained by varying the amount of hydrogen in the initial mixture. Because we need to keep the iron-to-oxygen ratio constant, our opacity tables for the ``new''-solar metallicity have reduced iron abundances. Therefore, our ``old''-solar models are too oxygen rich, whereas the ``new''-solar models are iron poor. While this is not accurate, it does mean that we expect the evolution of stars in the Galaxy to be between these two extremes. During helium burning, carbon and oxygen is rapidly synthesised replacing the initial abundances, and as a consequence, changing the initial oxygen abundances would manifest only slight differences. In addition during hydrogen burning, which we do not examine in this work, most of the material is converted to nitrogen via Carbon-Nitrogen-Oxygen-cycle processes. Finally, another difference between ``old''- and ``new''-solar is the different amounts of neon--with ``new'' solar being more neon rich. In our models, any excess nitrogen from the main sequence is rapidly burnt to neon-22 before helium burning begins in earnest. Thus, reducing the impact of starting our models as too oxygen rich.

\begin{table}
	\centering
	\caption{A comparison of compositions between our models and the work of \citet{2012A&A...539A.143N}.}	

	\begin{tabular}{l  l  l}\\
		\hline\hline
		Source	&	[Fe/H]	& [O/H] \\\hline
		Nieva \& Przybilla (2012) & $7.52\pm0.03$ &  $8.76\pm0.05$ \\
		$Z=0.020$, $X=0.700$  &  7.56   &  8.93   \\
		$Z=0.014$, $X=0.715$  &  7.40  & 8.70  \\
		$Z=0.008$, $X=0.730$  &  7.15  &  8.51   \\
		\hline
		\label{tab:ohfe}
	\end{tabular}
\end{table}

The primary effect of the metallicity important for this work is how the mass-loss rates scale the different metallicities and how iron contributes to the opacity in determining the inflation in our models. As we see from the values of [Fe/H] in Table~\ref{tab:ohfe}, we should choose to scale our mass-loss rates from our $Z=0.020$ models. We have also created a grid scaling from $Z=0.014$ and find only slight differences from shifting the scaling point. To further understand this we have looked at the abundances of the stars used in \citet{2000A&A...360..227N}, where we took the WR mass-loss rates applied to our models. Using the list of WR stars in the paper, we calculated the distance of each WR star from the Galactic centre, and then estimated the [Fe/H] and [O/H] abundance of each star using the local cosmic abundances and composition gradients given in \citet{2012A&A...539A.143N}. We find that the WR mass-loss rates of \citet{2000A&A...360..227N} are based on a sample of stars with values of [Fe/H] that range between 7.35 and 7.77, with a mean of 7.54$\pm$0.08. Whereas for [O/H] the range is between 8.65 and 8.93, with a mean of 8.77 $\pm$0.05.

Therefore, again, we have some confidence in our iron abundance being correct; however, as stated above, our initial oxygen abundance may be slightly too high. Nonetheless, with hydrogen-burning processes this oxygen will have been mostly converted to nitrogen, so the effect on the WR stars will be minimal.

\begin{figure}
	\centering
	\includegraphics[width=\columnwidth]{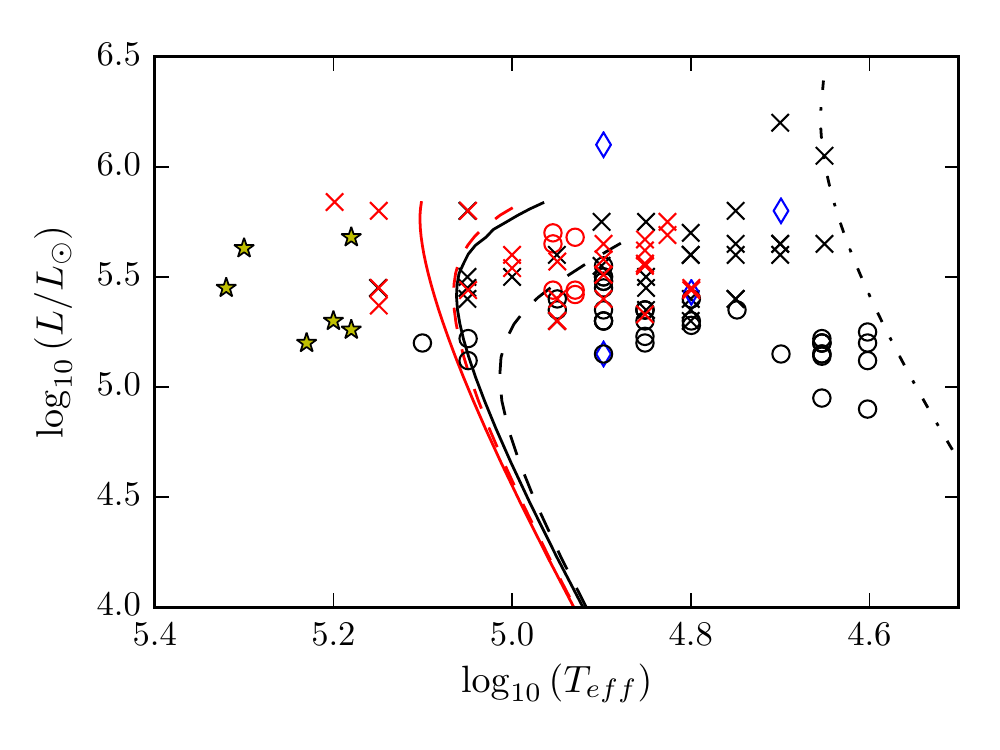}
	\caption[]{HR diagram showing the helium zero-age main sequence for models with and without clumping represented by solid and dashed lines, respectively. The colour of the line indicates the metallicity: $Z=0.008$, red; and $Z=0.02$, black. The location of the HZAMS, for $Z=0.02$, is indicated with a black, broken line. The observed WR locations are detailed in section \ref{ssec:obswrstars} and are marked as follows: WN, saltires; WC, circles; WO, yellow stars; and WN/WC transition objects, blue diamonds. Black represents Galactic observations, while red represents observations in the LMC.}
	\label{fig:HeZAMS}
\end{figure}

To summarise, we shall consider pure helium models at four different metallicities: $Z=0.008,0.014,0.02,0.04$; and with a helium mass fraction, $Y = 1.0 - Z$. With no hydrogen, the model immediately ignites helium in the core upon starting the evolution. The models are then evolved until neon ignition or the end of carbon burning. All models begin as chemically homogeneous stars on the helium zero-age main sequence (HeZAMS); the location of which is metallicity dependent as seen in Fig.~\ref{fig:HeZAMS} (see also section \ref{ssec:hezams}).  The size of the convective core is extended via a prescription of overshooting \citep{1997MNRAS.285..696S} with an overshooting parameter set at $\delta_{ov}=0.12$.

\subsection{Mass-loss scheme}
\label{ssec:mlscheme}

We employ a mass-loss scheme, outlined in \citet{2006A&A...452..295E}, that is derived from \citet{2000A&A...360..227N}. Various factors influence the mass-loss rate; however, most crucial is metallicity due to the effect it has on the opacity in the stellar envelope \citep{2005A&A...442..587V,2006A&A...452..295E}. Thus, due to stronger radiation driving on the metal lines, it is expected that metal-rich stars will exhibit a significant enhancement in envelope opacity, and therefore, an increased mass-loss rate. The mass loss employed in the evolutionary calculations is of the form,
\begin{equation}
	\dot{M'} \propto \dot{M} \beta \left( \frac{Z}{Z_{\odot}} \right)^{0.7},
\end{equation}
where $\dot{M}$ is taken from \citet{2006A&A...452..295E}, $Z$ is the heavy metal abundance of the model, and $Z_{\odot}$ is solar metallicity. Note that the choice of solar metallicity depends on the model metallicity: ``new'' solar metallicity is used for models with $Z=0.014$; whereas ``old'' solar metallicity is used for models with $Z=0.008, 0.02 \textrm{ and } 0.04$. For the low-luminosity helium stars, the mass-loss rates are extrapolated. However, there is some uncertainty about the validity of this \citep[see discussion in][]{2015PASA...32...15Y}. We note that, as we vary the mass-loss rates for stars below the typical Wolf--Rayet luminosities, the effect on the mass-loss evolution is small. It is here we introduce $\beta$, a scaling factor allowing for variation of the mass-loss rate.

The reason we use such a means for varying the mass-loss rate is two-fold. Firstly, it enables us to see how evolution is affected by changing the mass-loss rates, as they are uncertain. 
Secondly, it allows us to estimate what the evolution would be like if, before the hydrogen envelope were removed, more helium burning occurred. 
For example, in the case of $\beta=1$ the hydrogen envelope is removed before the beginning of helium burning, so the tracks represent the greatest possible effect of mass loss on the models. Thus, more helium mass is lost from the WR stars. 
However, for $\beta=0$ the evolution towards the end of the track represents how the star would appear if the hydrogen envelope were only removed near the end of helium burning. 
So, post-helium burning, the structure would show a more massive helium envelope than in the previous case. In effect, more massive stars will lose their hydrogen envelope quicker, and thus, evolve like the $\beta=1$ case. Conversely, less massive stars would only lose their hydrogen envelope after the completion of helium burning, and would look similar to the $\beta=0$ case (i.e. no mass loss). While models with no mass loss are unlikely to exist in reality, especially for massive stars, we include them for completeness and illustrative purposes. Effectively, these models indicate the maximum amount of inflation that may be possible without including a clumping factor to increase the opacity. We therefore only concern ourselves with the mass of the helium envelope.
We use this knowledge to admit ourselves a discussion on why having no mass loss ($\beta=0$) in our models gives results that seemingly represent the observed positions of WN stars, and to estimate how models with very weak mass-loss rates might evolve.

\subsection{Envelope inflation}

Massive stars develop an extended envelope structure due to an increase in opacity caused by the iron-opacity peak. At stellar temperatures near the iron-opacity peak ($\log{T_\mathrm{eff}} \sim 5.2$), a small convective layer forms above a near-void region of the star. The overall effect is a radiation-driven expansion of this high opacity outer layer; a reduction of the apparent stellar temperature. The effect is more severe with increasing metallicity due to the increased abundance of iron-group elements.

We allow the material in inflated envelope to be inhomogeneous and define a clumping factor, $D$, as described by \citet{2012A&A...538A..40G}. To investigate the effect of density inhomogeneities in the inflated envelopes of helium stars, we select two values for our clumping factor: $1$ and $10$ referring to ``unclumped'' and ``clumped'', respectively. 

When referring to a particular set of models, we shall label them by the metallicity, $\beta$, and clumping factor like so: ($Z$, $\beta$, $D$). For example, (0.02, 1, 1) refers to the grid of models with $Z=0.02$, $\beta=1$, and a normal envelope without the clumping factor to enhance the opacity.

All the models we have computed are freely available online. They can be found at the \textsc{bpass} (Binary Population and Spectral Synthesis) code website, \texttt{http://bpass.auckland.ac.nz}.

\begin{figure}
	\centering
	\includegraphics[width=\columnwidth]{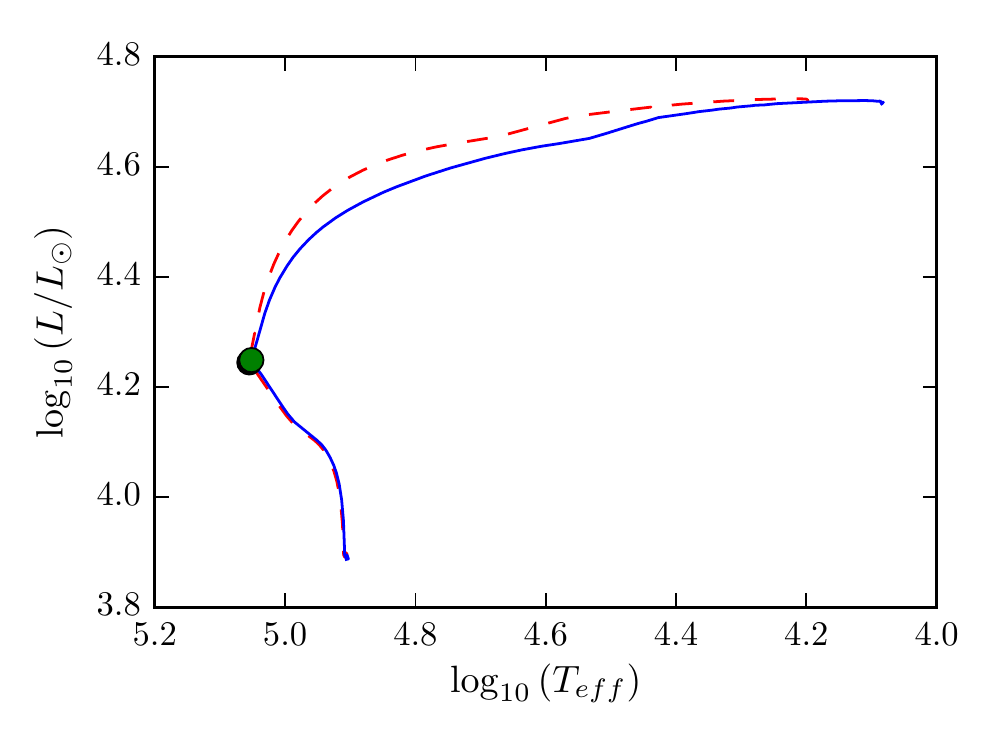}
	\caption[]{HR diagram showing the evolutionary tracks (see text) of a $3\msun$ helium star at solar metallicity with (solid, blue) and without (broken, red) clumping. The ignition point of the helium shell is marked on each track.}
	\label{fig:3MBurningPhases}
\end{figure}

\section{Results}
\label{sec:results}

\subsection{Location of the HeZAMS}
\label{ssec:hezams}

We can see in Fig.~\ref{fig:HeZAMS} the theoretical location on the HR diagram of the helium zero-age main sequence for two of the metallicities considered--$Z=0.008$ and $Z=0.02$. For initial helium star masses $\lesssim13\,\msun$, the HeZAMS location is mostly unaffected by a change in metallicity. However, for larger initial stellar masses, the HeZAMS location shows a strong dependence on metallicity, bending towards cooler stellar temperatures with increasing metallicity \citep[cf.][Fig. 1]{2006A&A...450..219P}.

The cause of this bend is inflation of the stellar envelope \citep[see][for details]{2012A&A...538A..40G,2006A&A...450..219P}. The inclusion of clumping further intensifies the bending.

Plotted alongside the HeZAMS are the locations of the Galactic WN stars modelled by the Potsdam WR group \citep{2006A&A...457.1015H}. We see that most of observed WN stars lie within a region far cooler than that of the HeZAMS. Thus, even though higher metallicity favours cooler stellar temperatures for massive helium stars, the bending of the HeZAMS is not enough to reproduce most of the observed WN locations. We can, therefore, only assume that either our stellar models lack some physical details, or the atmosphere models do. However, all stellar evolution models have the same mismatch, and both the Potsdam models and the \textsc{CMFGEN} models provide similar results for WR stars \citep{2012A&A...540A.144S}.

\subsection{Evolution of low-mass helium stars}
\label{ssec:lowmassheevo}

We broadly distinguish the evolutionary behaviour of helium stars by defining two classes of model: those with low initial mass, and those with high initial mass.

We may generalise the evolution of low-mass helium stars as eventually experiencing a phase of evolution as a helium giant--similar to the red-giant phase of post-main-sequence hydrogen stars. We show the evolution of a typical the $3\msun$ star in Fig.~\ref{fig:3MBurningPhases}. The stellar luminosity increases along the helium main sequence until core helium is exhausted. An important consequence of the low mass-loss rate is that, as evolution proceeds, the helium envelope is retained. After the helium shell ignites, the envelope expands and the stellar temperature decreases.

\begin{figure*}
	\centering
	\includegraphics[angle=0,width=0.33\textwidth]{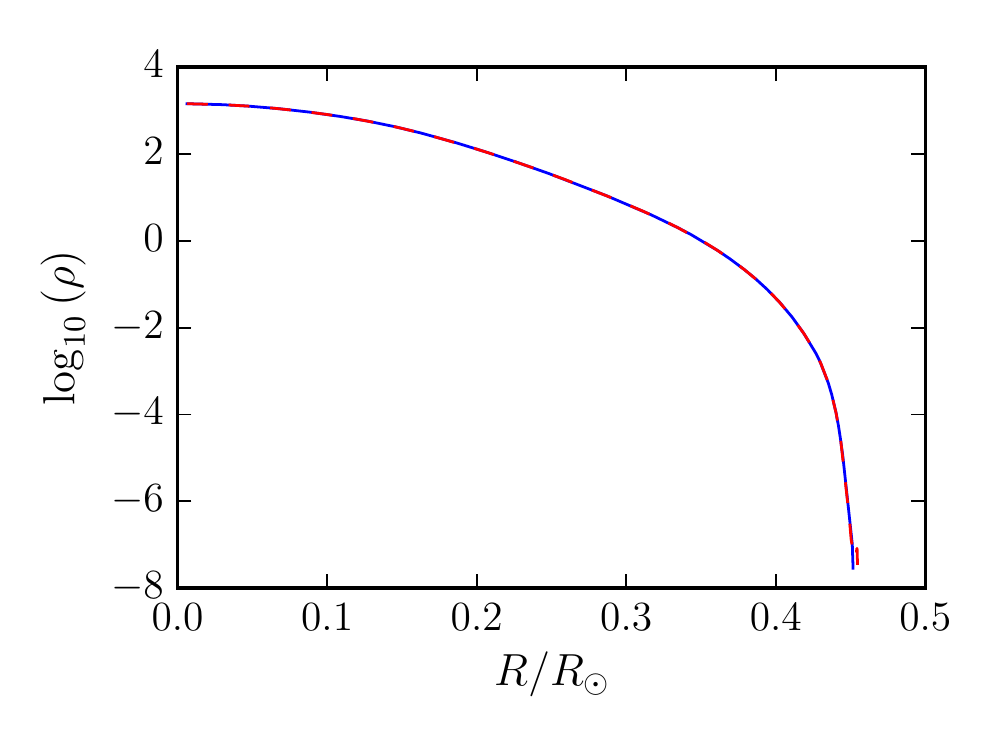}
	\includegraphics[angle=0,width=0.33\textwidth]{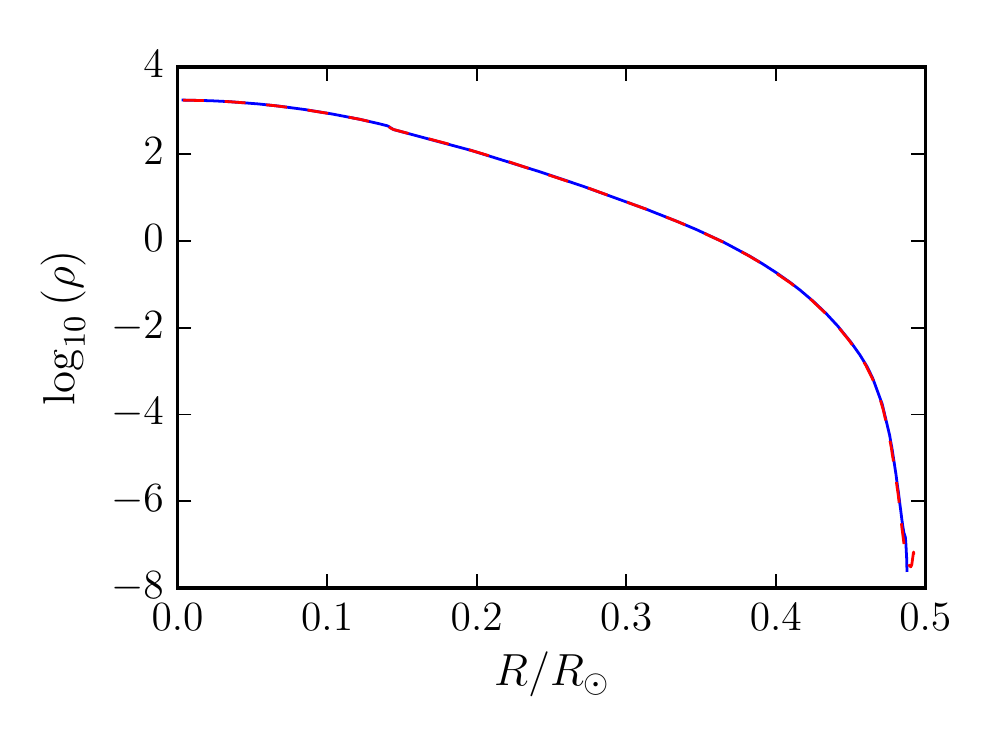}
	\includegraphics[angle=0,width=0.33\textwidth]{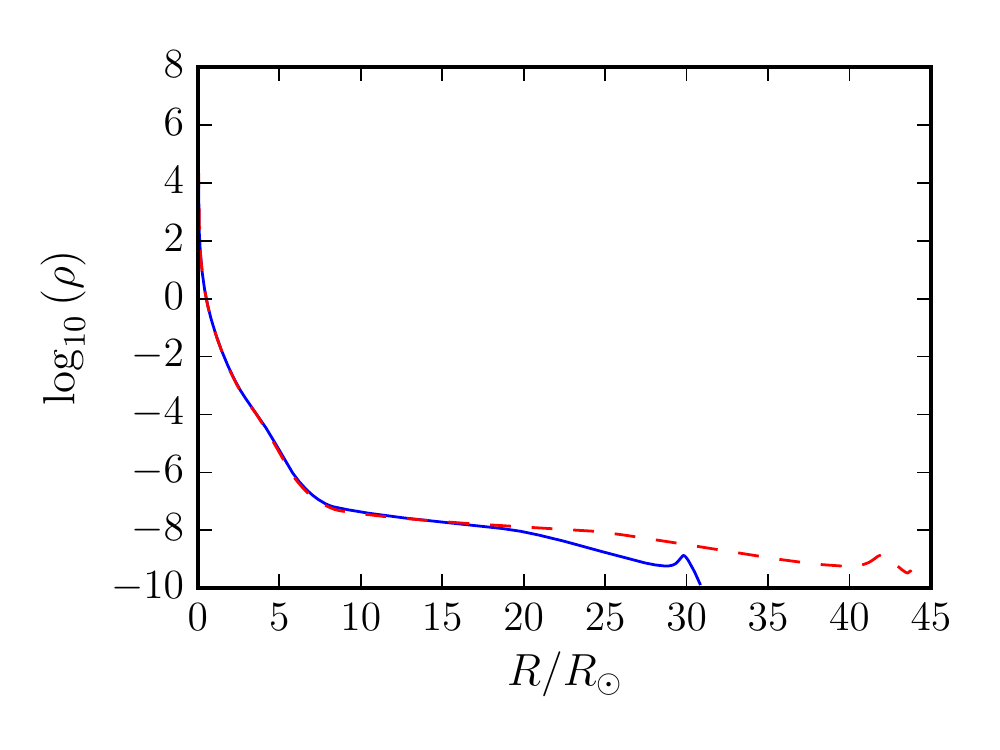}
	\caption[]{Density profile of a $3\,\msun$ helium star at various evolutionary stages. Left, on the HeZAMS; middle, at the onset of helium-shell burning (core-helium exhaustion); right, at model termination. The solid, blue line is for $D=1$; the broken, red line is for $D=10$. Models are evolved with $\beta=1$.}
	\label{fig:densHGlifetime}	
\end{figure*}

\begin{figure*}
	\centering
	\includegraphics[angle=0,width=0.33\textwidth]{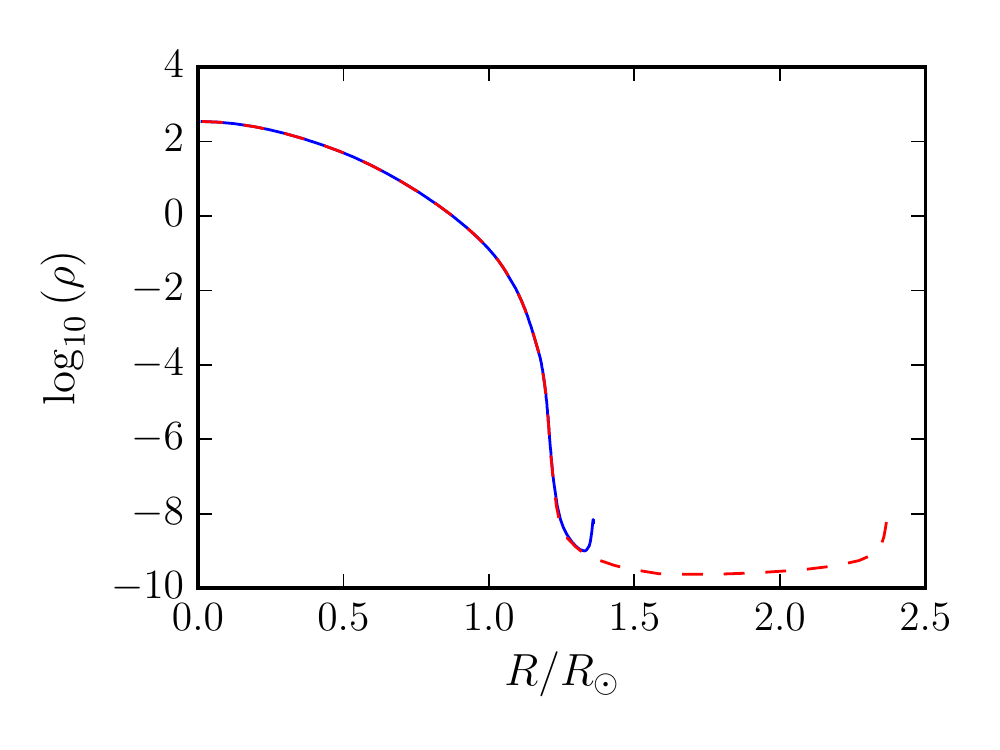}
	\includegraphics[angle=0,width=0.33\textwidth]{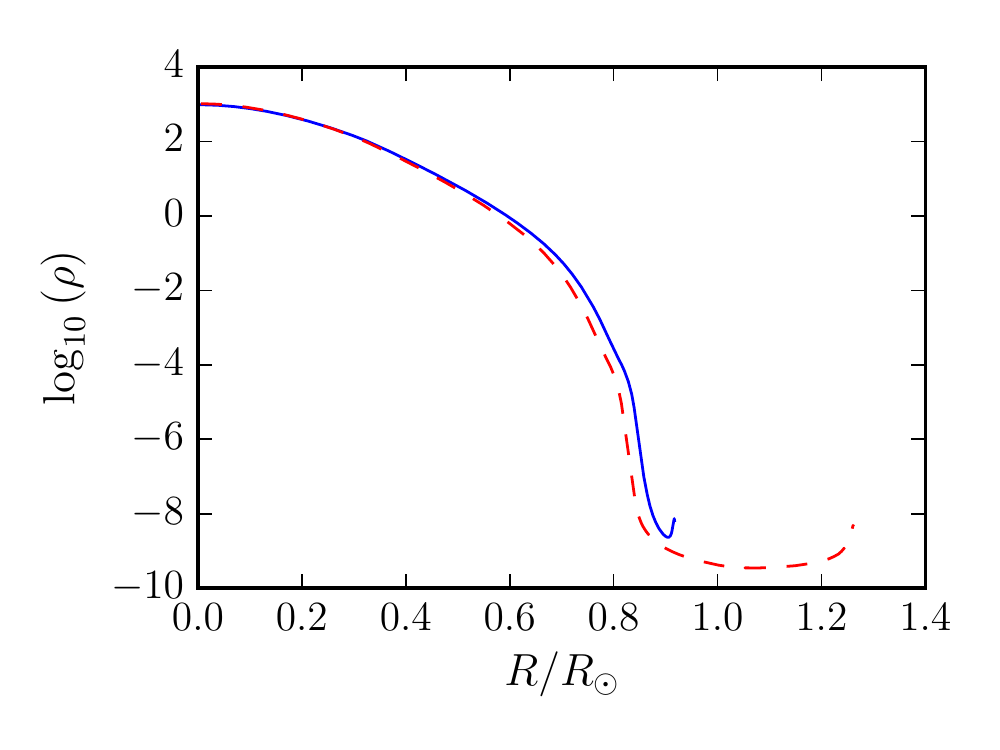}
	\includegraphics[angle=0,width=0.33\textwidth]{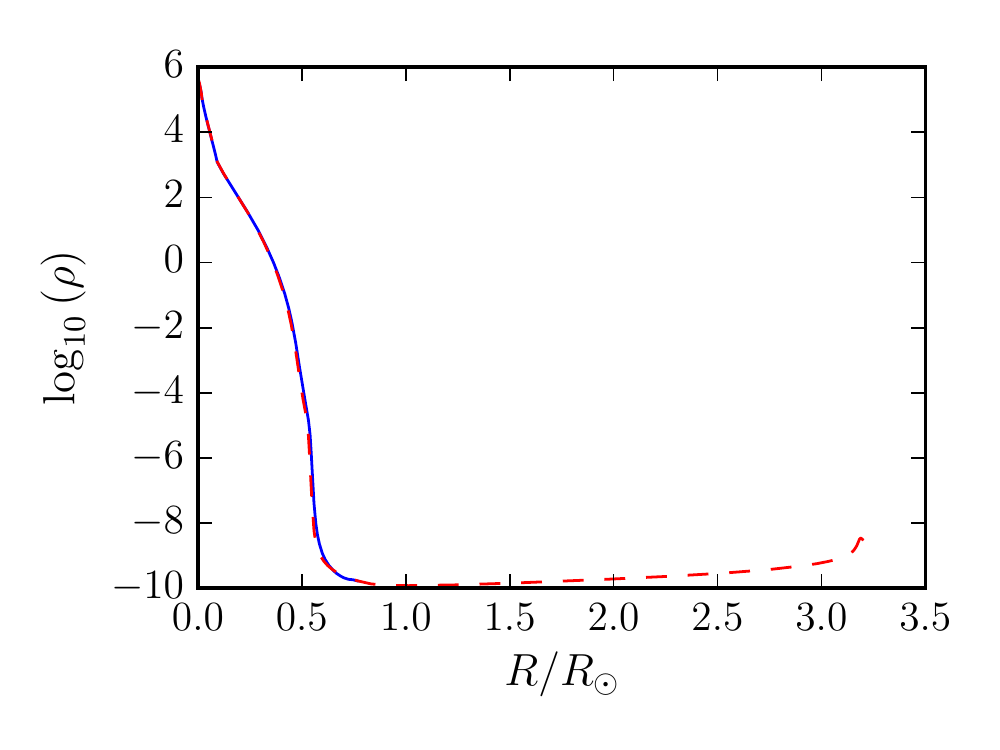}
	\caption[]{Density profile of a $15\,\msun$ helium star at various evolutionary stages. Left, on the HeZAMS; middle, at the onset of helium-shell burning (core-helium exhaustion); right, at model termination. The lines have the same meanings as in Fig.~\ref{fig:densHGlifetime}}
	\label{fig:densWRlifetime}	
\end{figure*}

Fig.~\ref{fig:densHGlifetime} shows the structure of a typical low-mass helium star model (here, $3\msun$) at various stages of evolution. At the HeZAMS,  the density profile is very simple: a dense central region that decreases in density towards the surface. The evolution through core-helium burning proceeds as expected: the consumption of helium in the core causes the model to contract, increasing the central density. After the start of shell-helium burning, the envelope expands and the star cools. Following shell burning, the structure of the model is similar to that of a red-giant star: a dense core region with an expansive envelope. The envelope of this ``helium giant'' is punctuated by a small density inversion at the surface (cf. Fig.~\ref{fig:densWRlifetime}). The low luminosity of the model--in comparison to that of a WR star--implies a much reduced mass-loss rate causing the star to retain its helium-rich outer envelope for the entirety of its evolution. As a result, the rate of mass loss is insufficient to strip off the outer layers of the star, and instead, expands the stellar envelope through the helium-shell-burning phase. 

Our findings indicate that a helium-giant phase is characteristic of helium stars with initial masses of up to $\approx 8\msun$ (for $Z=0.02$) with higher initial masses being categorically WR-type stars. This represents an upwards revision to the initial mass threshold of WR-type evolution for helium stars \citep[cf.][with $\approx 5\msun$]{2002PASA...19..233P}.
The reason for this increase is most likely due to the newer opacity files employed from \citet{2004MNRAS.348..201E}, which more accurately follow the changes in opacity as the carbon and oxygen abundances vary.
It should be noted that the threshold mass for a helium-giant phase decreases with metallicity because of the decrease in envelope opacity (discussed below). For $Z=0.008$, a helium-giant phase is present for initial masses up to $\approx 5\msun$.

The low-mass helium models are, as expected, of low luminosity. As the mass-loss rates we use are related to the luminosity, we find the final masses of these models are close to their initial masses. Indeed, as may be seen in Figs.~\ref{fig:WRObs} and \ref{fig:WRObsZ008}, the low-mass models remain at near-constant luminosity. From these figures we note that, regardless of our choice of $\beta$, the evolution for low-mass helium models is relatively similar. Although the $1\msun$ model does experience a helium-giant phase before the model progresses towards becoming a white dwarf, models with a greater initial mass (i.e. $>2\msun$) evolve to explode as SNe.

In Fig.~\ref{fig:density3Mand15M}, we compare the structure of similar models at different metallicities and clumping factors. The higher metallicity models have greater opacities, so to maintain equilibrium, the models must have greater radii. As a result, the radii of metal-rich models are comparatively larger than the radii of metal-poor models. In addition, the inclusion of clumping effects an increase in the radius of the model, which results in a slight 0.1 dex decrease in the surface temperature.

\setcounter{figure}{8}
\begin{figure*}
	\centering
	\includegraphics[angle=0,width=\columnwidth]{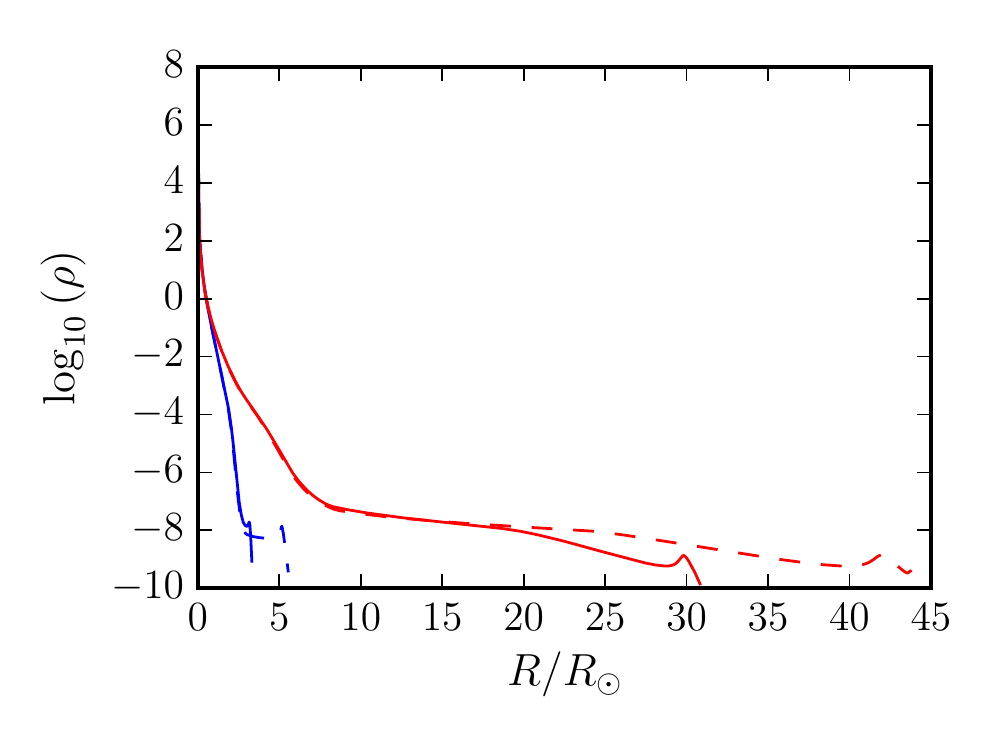}	
	\includegraphics[angle=0,width=\columnwidth]{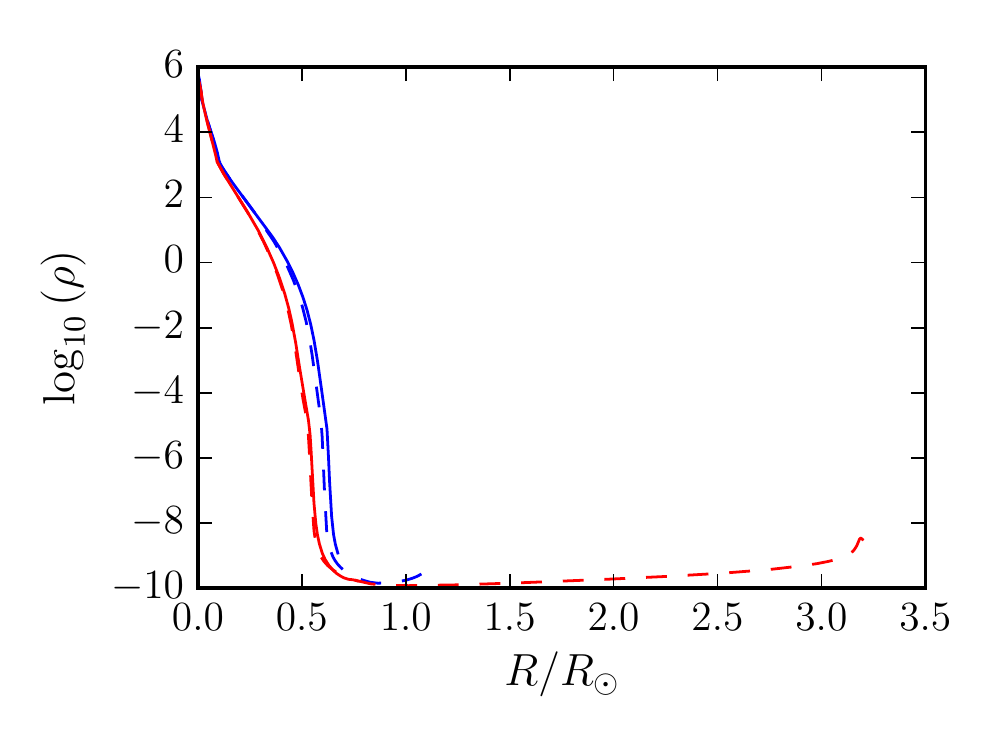}	
	\caption[]{Left: Density profile of a $3\, \msun$ helium star model at the termination of evolution. The blue and red lines denote $Z=0.008$ and $Z=0.02$, respectively. Additionally, the solid and broken lines denote a clumping factor of $D=1$ and $D=10$, respectively. Right: The same for a model of mass $15\, \msun$. In all cases, $\beta=1$.}
	\label{fig:density3Mand15M}
\end{figure*}
\subsection{Evolution of high-mass helium stars}
\label{ssec:evohigh}

In section~\ref{ssec:lowmassheevo} we discussed the evolution of helium star models with initial masses below $\approx 8\msun$. Here we consider the evolution of models above this threshold: models with WR-type evolution. WR-type evolution is characterised by high temperatures due to strong mass loss \citep{2007ARA&A..45..177C}. In Fig.~\ref{fig:WRBurningPhases}, we show the evolution of a $14\msun$ helium star model. By comparing the luminosities of our models with those of the observed WR stars, we see that to be a WR star a model must have $\log(L/L_{\odot}) \ga 4.9$.

During core-helium burning, a high-mass helium star experiences sharp mass loss, and thus, a decrease in mass. From the mass-luminosity relation, a reduction in mass will decrease the luminosity of the star. As evidenced by Figs.~\ref{fig:WRObs} and \ref{fig:WRObsZ008}, models with higher mass loss experience a sharper drop in luminosity during the core-helium-burning phase; the drop in luminosity is, naturally, enhanced by an increased metallicity. Conversely, the absence of mass loss entirely affects the temperature range of the post-helium-burning phases. The effect is most notable with (0.02, 0, 1), where the model progresses to a lower effective temperature as opposed to a ``loop'' that is characteristic of the Wolf--Rayet phase. With the removal of the outer layers of helium, the envelope expands during the shell burning of helium. 

The difference in structure is immediately apparent by comparing Fig.~\ref{fig:densWRlifetime} with Fig.~\ref{fig:densHGlifetime}. Whereas a low-mass model begins core-helium burning with a dense core region that decreases in density towards the surface, a high-mass model has an extended region of near-constant density that ends with a large density inversion (a ``bump'') at the surface. As mentioned in section~\ref{ssec:hezams}, the density inversion sits atop the extended envelope structure of the high-mass helium star models, and because of the relationship with the iron-opacity peak, is affected by metallicity.

This surface density inversion is removed during core-helium burning by strong mass loss, as evident in Fig.~\ref{fig:densWRlifetime} (middle). At this point in the evolution, the envelope is completely stripped, which leaves the core is fully exposed. 
When we include clumping in the models, the radial expansion is predictably enhanced. A notable difference to the unclumped model is that the mass-loss rate is now insufficient to remove the surface density inversion.

Without mass loss ($\beta=0$), the effect of the expansion is more significant. We observe in Figs.~\ref{fig:WRObs} and \ref{fig:WRObsZ008} that higher mass helium star models reach comparatively cooler surface temperatures \citep{1999PASJ...51..417I,2006A&A...450..219P}. The amount of inflation gives an indication as to the size of the helium envelope in relation to the CO core. When $\beta>0$, the inflation is less substantial as more of the helium envelope has been removed. Therefore, we may infer information about the internal structure of a WR star from the surface temperature.
As noted previously, clumping enhances the effect of inflation in the stellar envelope. We can clearly see the result in  Figs.~\ref{fig:WRObs} and \ref{fig:WRObsZ008}. Clumped models evolve to reach far cooler stellar temperatures than unclumped model sets. As expected, the inflation of the stellar envelope is further enhanced by a higher metallicity.

\subsection{Populations of helium stars}
\label{ssec:obswrstars}

Figs. \ref{fig:WRObs} and \ref{fig:WRObsZ008} present our models with observed WR stars on the Hertzsprung--Russell diagram. Observational data is taken from \citet{2006A&A...457.1015H,2012A&A...540A.144S}, for Galactic WN and WC stars; \citet{2014A&A...565A..27H}, for Large Magellanic Cloud (LMC) WN stars; and \citet{2013A&A...559A..72T,2015IAUS..307..144T}, for WO stars. Also included are results from \citet{2002A&A...392..653C} for LMC WC stars. In this work, we attempt to reproduce the observed locations of WR stars. 

\subsubsection{Galactic WN stars}
\label{ssec:GalacticWN}
We begin our analysis by comparing the observed Galactic WN positions
with the theoretical WN positions (depicted with solid, green
lines). The observed early-type WN stars lie near the HeZAMS for
massive helium stars and are, generally, in good agreement with helium
star models of initial masses above $\approx 10\msun$. However, the
agreement is not so favourable for the observed late-type WN
stars. These WN stars have stellar temperatures that are far cooler
than those at the HeZAMS, and their locations cannot be reproduced by
using models with $\beta>0$.

Without mass loss ($\beta=0$), we see an interesting result: the
higher mass helium star models do, indeed, cross the region of
observed (hydrogen-free) late-type WN stars for solar metallicity
(``old'' and ``new''). A small amount of mass loss will remove the
outer layers of the envelope and expose the hot interior of the model;
thus, without mass loss, the model swells due to inflation and the
surface temperature decreases. The inclusion of envelope clumping
results in cooler stellar temperatures, as expected. However, even
with the inclusion of clumping, our helium star models are unable to
reproduce the observed locations of the coolest Galactic WN stars if
we use the standard WR mass-loss rates (i.e. $\beta=1$).

In section~\ref{ssec:mlscheme} we discussed the implications of
mass-loss rate (in effect, the parameter $\beta$) on the composition
at core-helium ignition. The value of $\beta$ relates to the size of
the helium envelope: a smaller value for $\beta$ will result in a more
massive helium envelope, and consequently, a lower surface
temperature.  Thus, models with $\beta=0$ are likely to be observed as
the coolest WN stars, while models with $\beta=1$ are the hottest WN
stars that reside near the HeZAMS. It is, therefore, apparent that
$\beta$ imposes constraints on the internal structure of WR stars; we
can use $\beta$ as an indicator of mass-loss history and helium
envelope mass.

To summarise, the large radii of the WN stars is due to either a small
amount of mass loss or clumping in the envelope.

\subsubsection{WN/WC transition object stars}

In our sample we also find four objects transitioning between the WN and WC phases. We note these are the most luminous, and thus, most massive stars in our sample. We find from our tracks that one, WR145, does agree with our reduced mass-loss models (with inflation). Two other stars, WR58 and WR126, appear in the same location on the HR diagram as other WN and WC stars. This suggests that they are, indeed, objects transitioning between the two types. The last star, WR26, is very luminous and we find that it is closest to our models at LMC metallicity. Therefore, WR26 likely has a mass beyond the upper limit of our grid--an initial mass of 25M$_{\odot}$.  We note that, for all of these transition objects, clumped models are required to achieve the observed locations on the HR diagram.  

\subsubsection{WC and WO stars}

The expected locations of WC stars (marked in solid blue lines) are in very poor agreement with the positions of observed WC stars. The standard evolutionary picture of WR evolution--suggested by \citet{2012A&A...540A.144S} where the WC phase succeeds a WN phase--is clearly insufficient to explain this discrepancy. We note from Fig.~\ref{fig:WRObs} that low-mass helium models can reproduce the observed locations of early- and late-type WC stars.

We note the observed WO stars on the HR diagram are hotter and more
luminous than the observed WC stars. A higher luminosity implies a
higher stellar mass, and we indeed find the observed WO stars in a
region predicted by our high-mass models. Though difficult to draw
definitive conclusions due to the lack of observational evidence, we
argue the standard description of WC stars used in evolutionary models
is incorrect and actually applies instead to WO stars.

As can be seen in Fig.~\ref{fig:WRObs}, the locations of the observed
WC stars coincides with the locations of the low-mass helium
giants. The surface composition identifies these low-mass helium
giants as WN stars, not WC stars. Due to the weak mass-loss rates of
the low-mass helium stars, nitrogen remains abundant on the stellar
surface rather than being removed, as is the case for the higher mass
stars. It is possible that the surface nitrogen may be removed by an
alternative mechanism of extra mixing \citep{2013ApJ...773L...7F}:
nitrogen is mixed into the stellar interior and removed via nuclear
processing.  We have tested this hypothesis by adding a small amount
of extra mixing in the radiative zones of our models. We find that it
is possible to decrease the nitrogen abundance and increase the carbon
and oxygen abundance without affecting the evolutionary tracks to a
significant degree.  Furthermore, models evolved from the pre-helium
main sequence may have different composition profiles that we have not
taken account of here. Therefore, similar models from the hydrogen
zero-age main sequence may exhibit WC-type compositions.

\begin{figure*}
	\centering
	\includegraphics[width=0.49\textwidth]{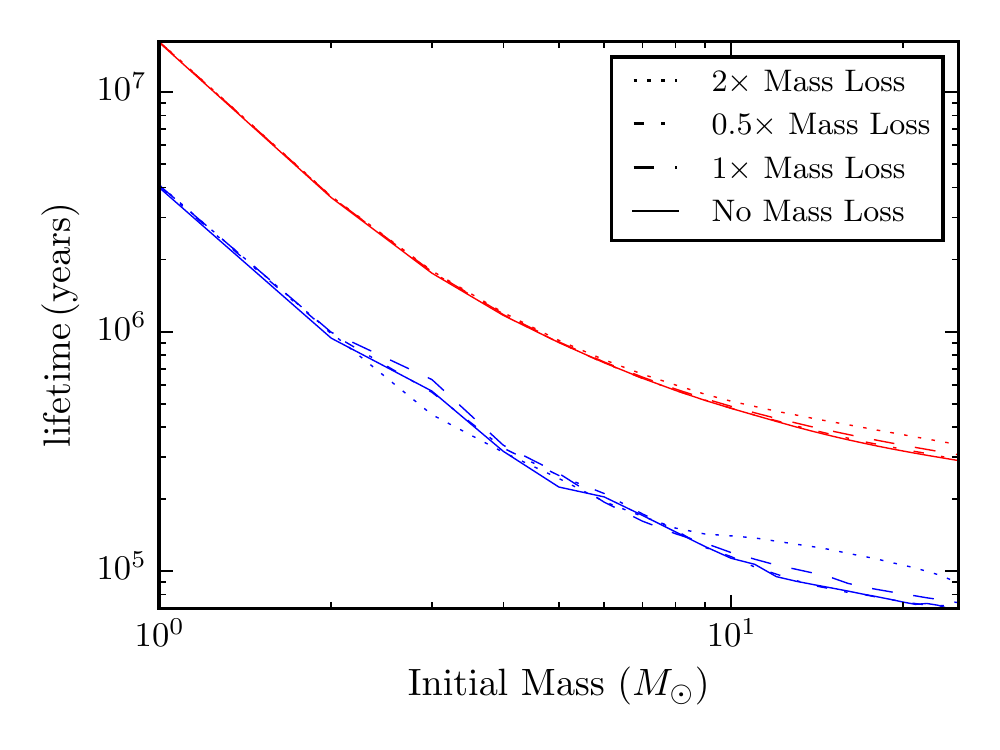}
	\includegraphics[width=0.49\textwidth]{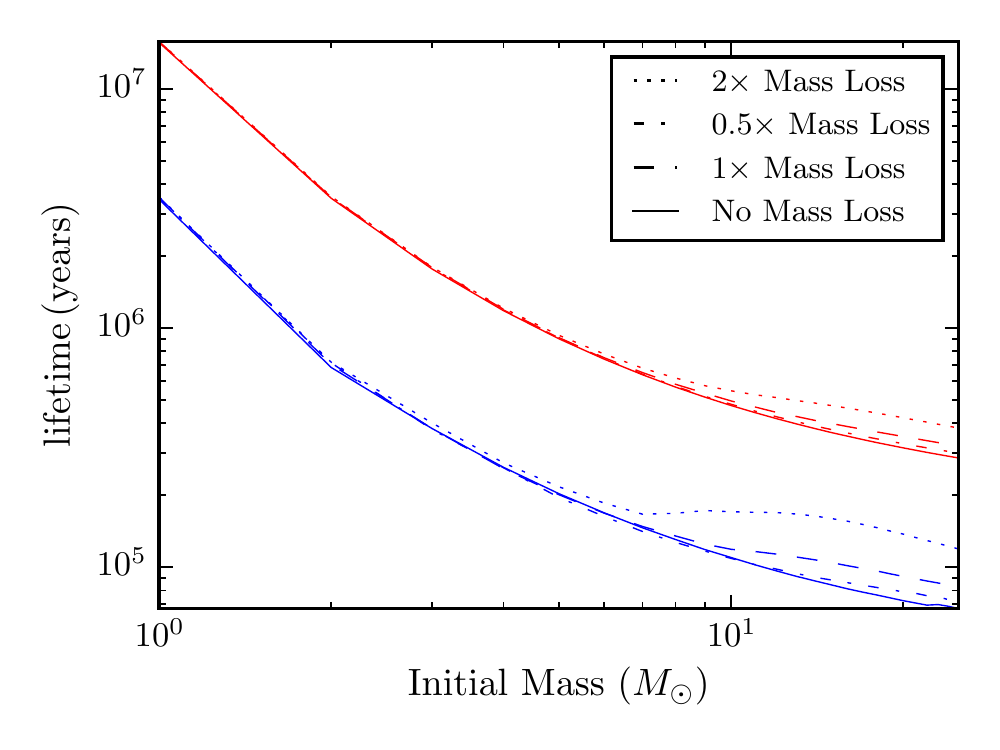}\\	
	\includegraphics[width=0.49\textwidth]{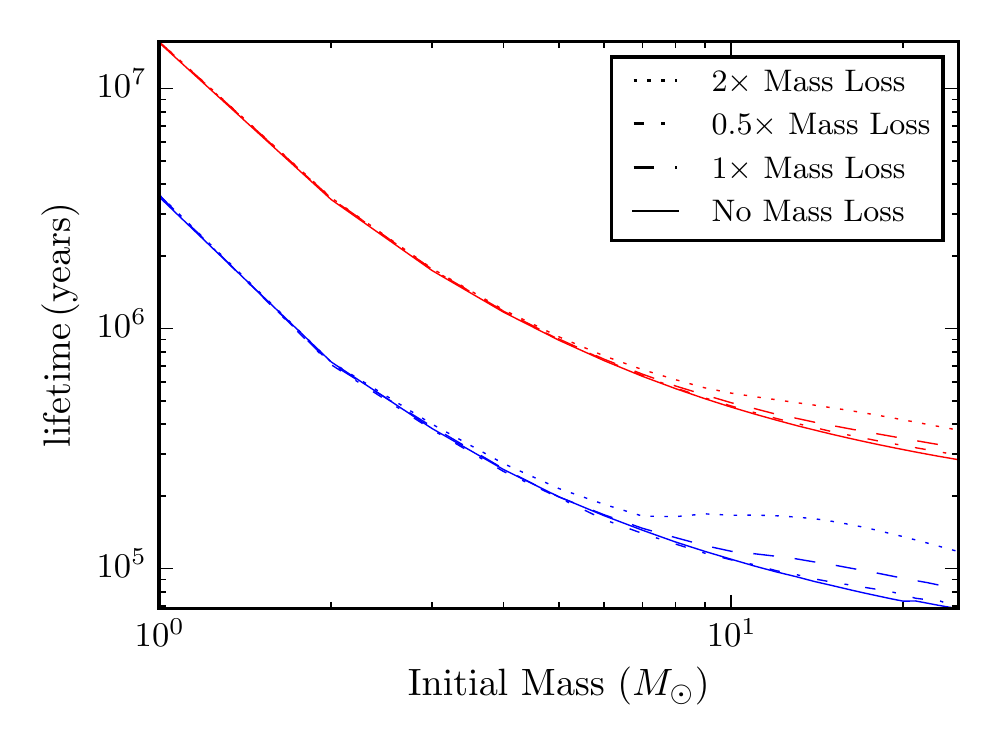}	
	\includegraphics[width=0.49\textwidth]{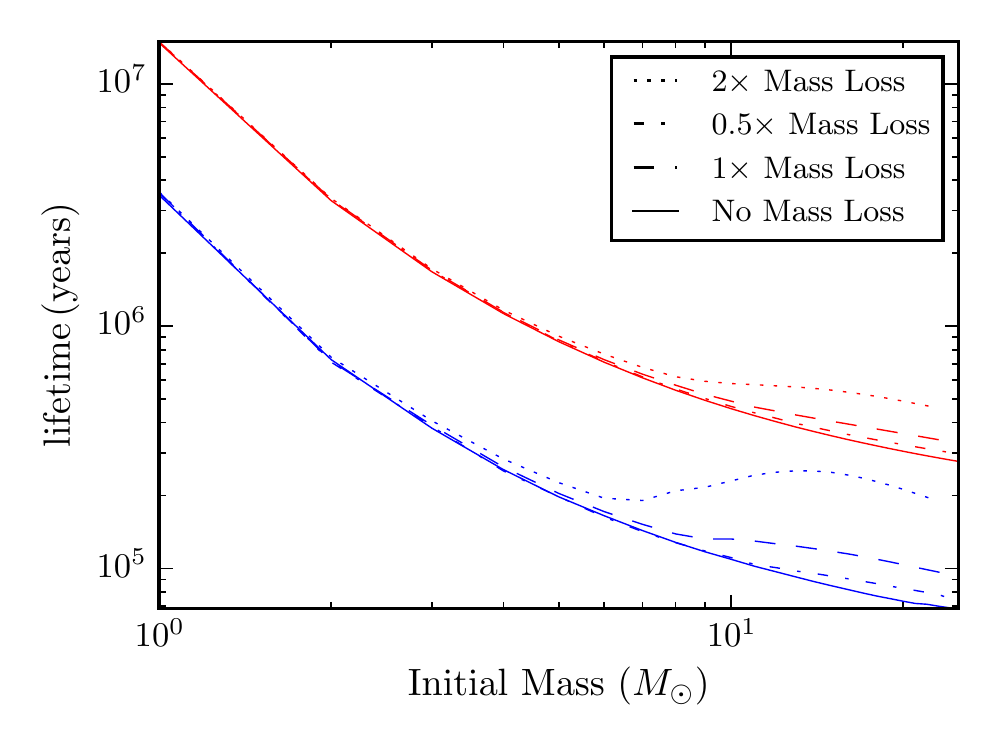}	
	\caption[]{Lifetimes of core-helium-burning phase (red) and post-helium-core-burning phases (blue) as a function of initial mass at different metallicities. Clockwise from the top-left: $Z=0.008$, $Z=0.014$, $Z=0.04$, and $Z=0.02$.}
	\label{fig:lifetimes_ML}
\end{figure*}

We note that, by including clumping in our stellar models, our models evolve closer to locations on the HR diagram where WC stars are observed. However, the coolest WC stars still arise from the same low-mass helium giants.

A further piece of evidence that this interpretation is correct can be
seen in Fig.~\ref{fig:lifetimes_ML}. If the WC stars do come from less
massive stars than the WN and WO stars, they will identify as WC only
after the completion of core-helium burning. In
Fig.~\ref{fig:lifetimes_ML}, we see that the post-helium-burning
lifetimes of the models for stars of initial masses $<8\msun$ are of
the same magnitude as the helium-burning lifetimes of the more massive
stars, which would be identified as WN stars. Furthermore, WO
stars--which arise from the most massive stars--have much shorter
lifetimes than the stars that we suggest are WC stars. This explains
why WO stars make up only a small fraction of the WR population.

In summary, our results suggest that both WC and WO stars are represented by models with $\beta=1$. However, they differ in terms of mass: WO stars are from the most massive of stars (an initial helium star mass of $\ga13\msun$), while WC stars come from less massive stars (an initial helium star mass of $\la13\msun$). Similar findings were made by \citet{2013A&A...558A.131G} and \citet{2012A&A...540A.144S}. We note that adding clumping improves the agreement but we find that WO stars must be unclumped.

\subsubsection{LMC WN stars}

In Fig.~\ref{fig:WRObsZ008}, we compare the observed LMC WR stars with
our models ($Z=0.008$). We expect, due to the low metallicity of the
LMC, a lower prevalence of inflated stars, and that WN stars in the
LMC will have higher stellar temperatures compared with Galactic WN
stars. We find that models without clumping (i.e. $D=1$) reproduce the
observed locations of LMC WN stars poorly.
\begin{figure*}
	\centering
	\includegraphics[width=\columnwidth]{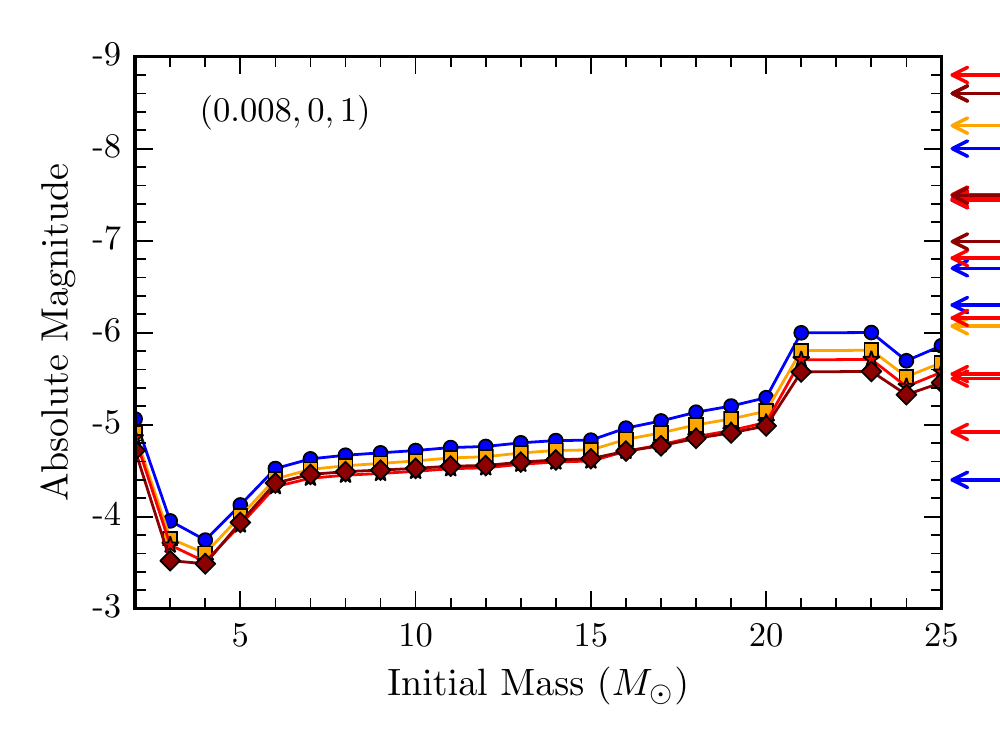}
	\includegraphics[width=\columnwidth]{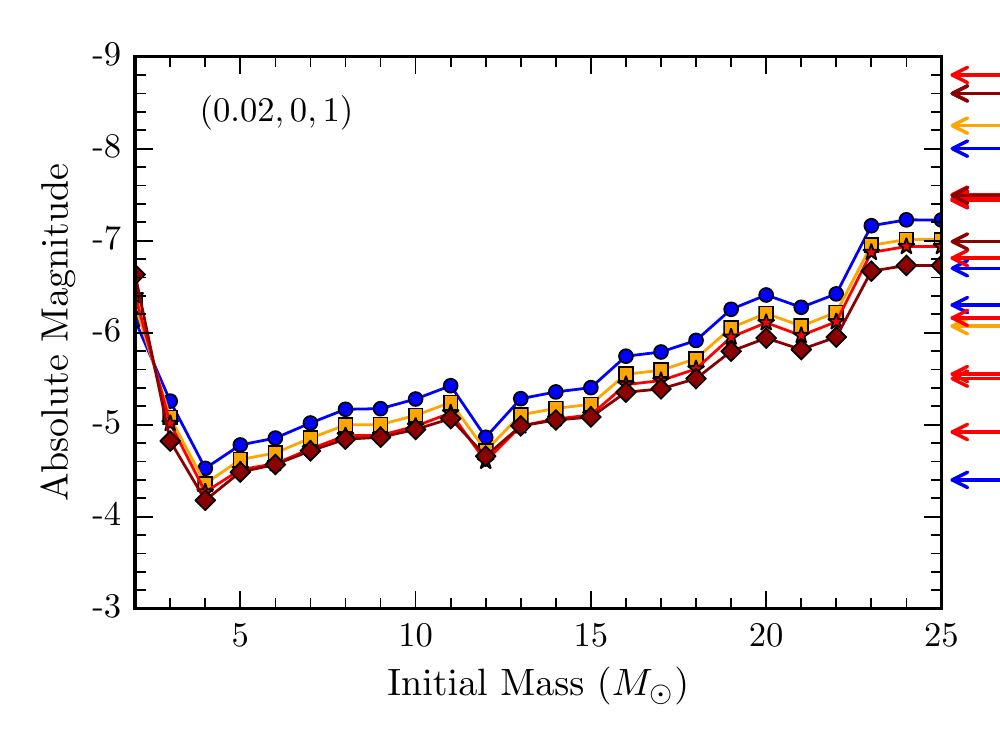}\\
	\includegraphics[width=\columnwidth]{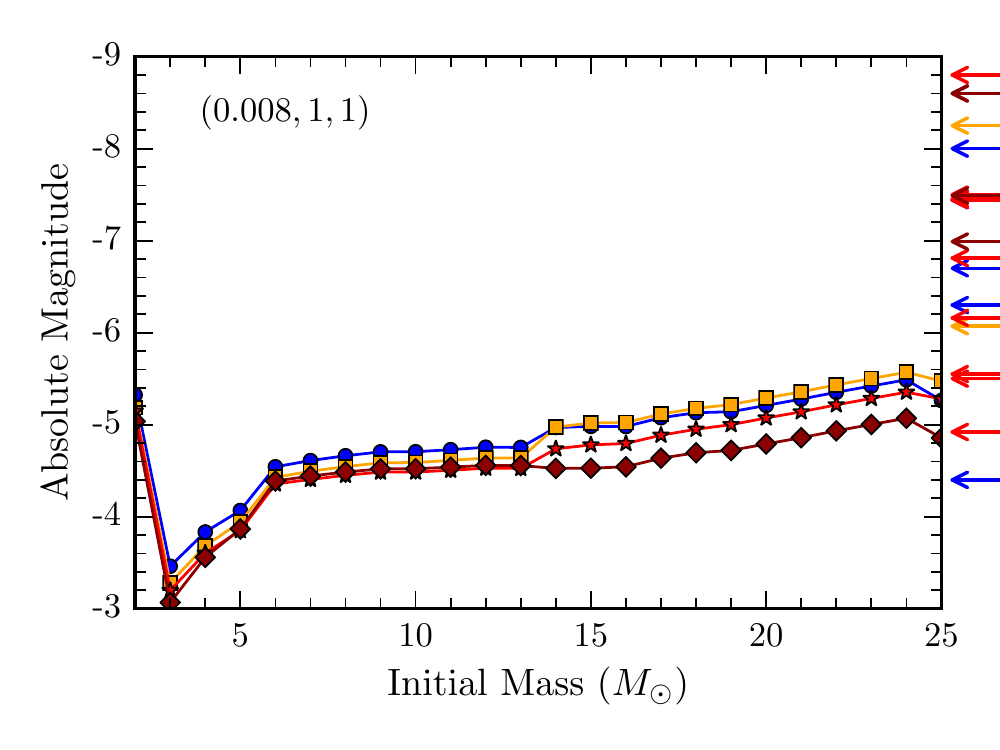}
	\includegraphics[width=\columnwidth]{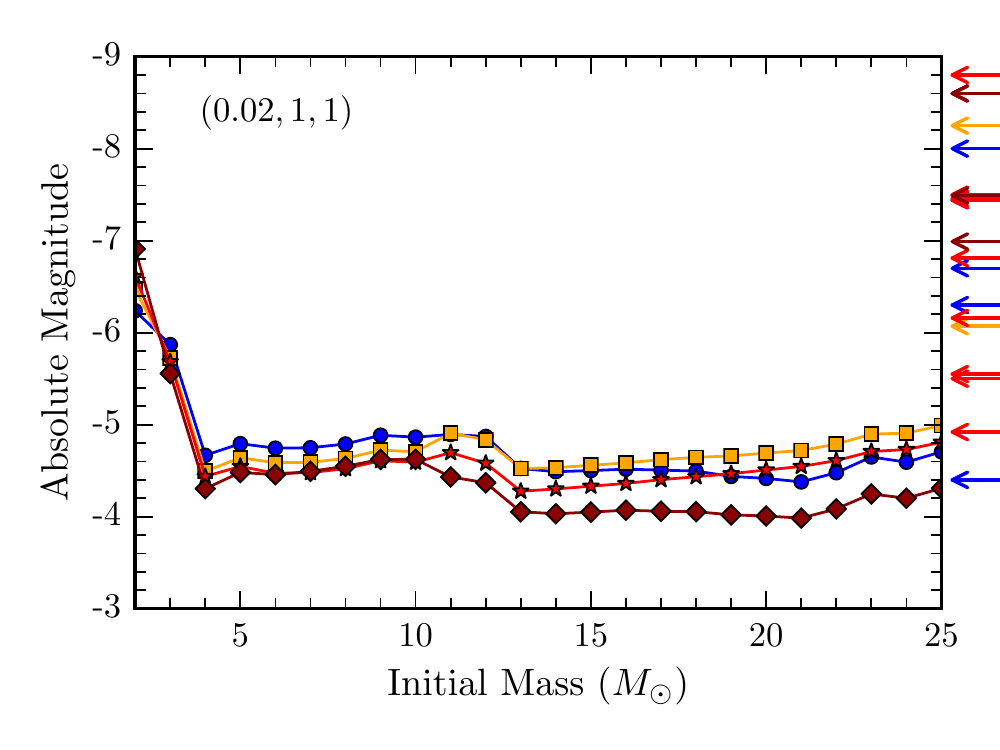}		
	\caption[]{Predicted supernova magnitudes of helium stars for models with $D=1$. The arrows on the right ordinate denote the limits of observed progenitors \citet{2013MNRAS.436..774E}; filters B, V, R and I are represented by colours blue, yellow, bright red and dark red respectively.}
	\label{fig:SNeMagsNoInflation}
\end{figure*}
\begin{figure*}
	\centering
	\includegraphics[width=\columnwidth]{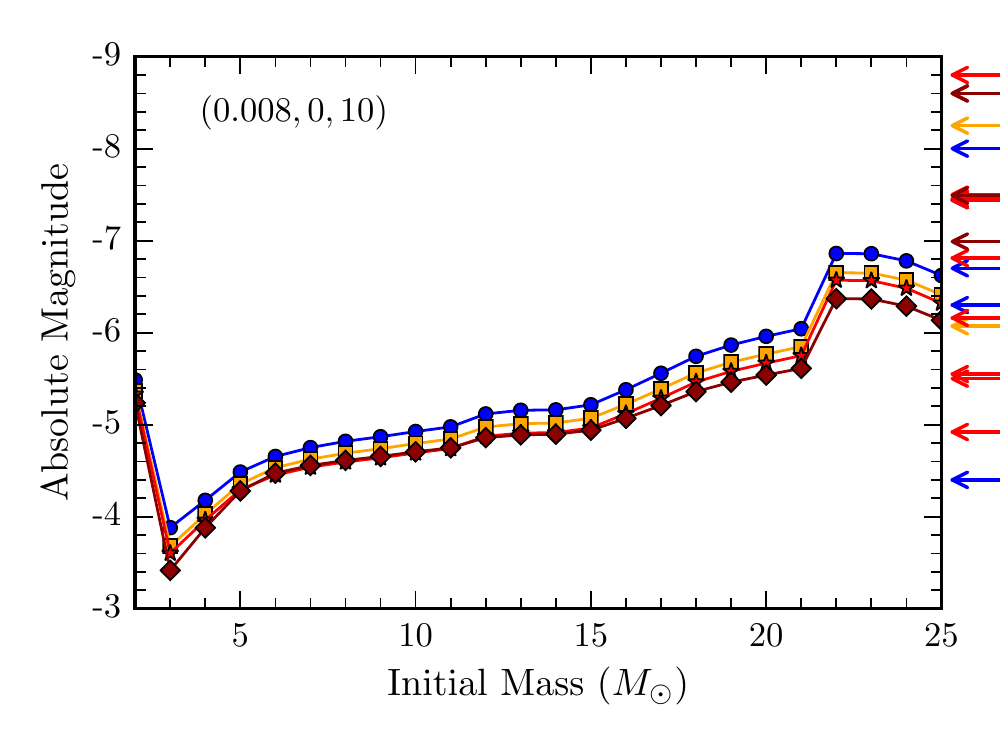}
	\includegraphics[width=\columnwidth]{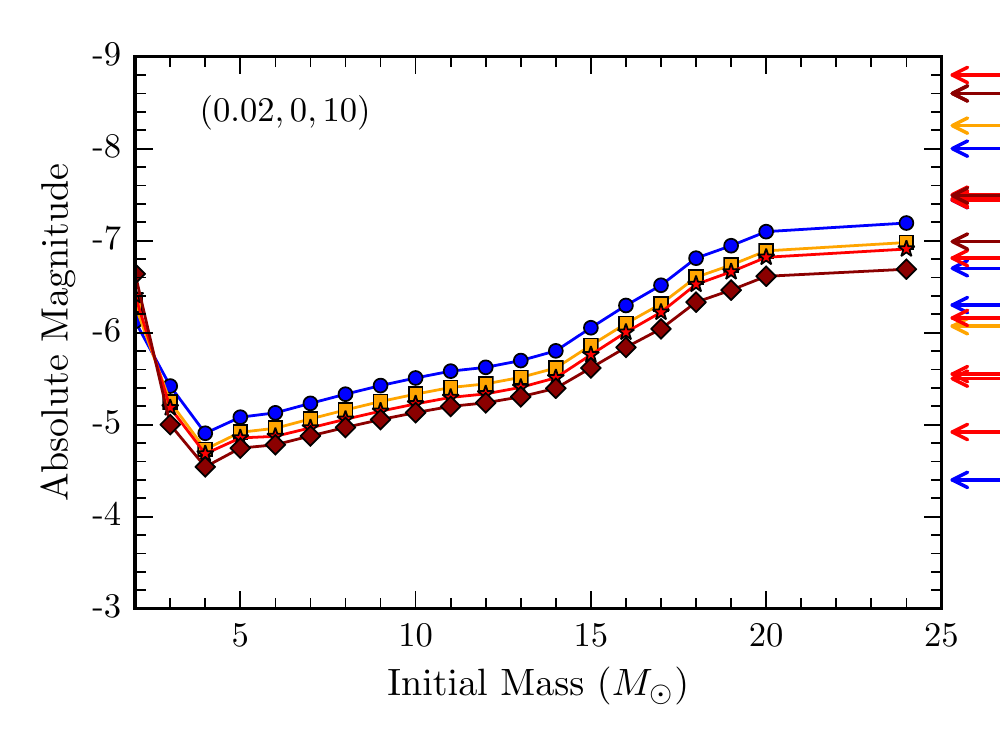}\\
	\includegraphics[width=\columnwidth]{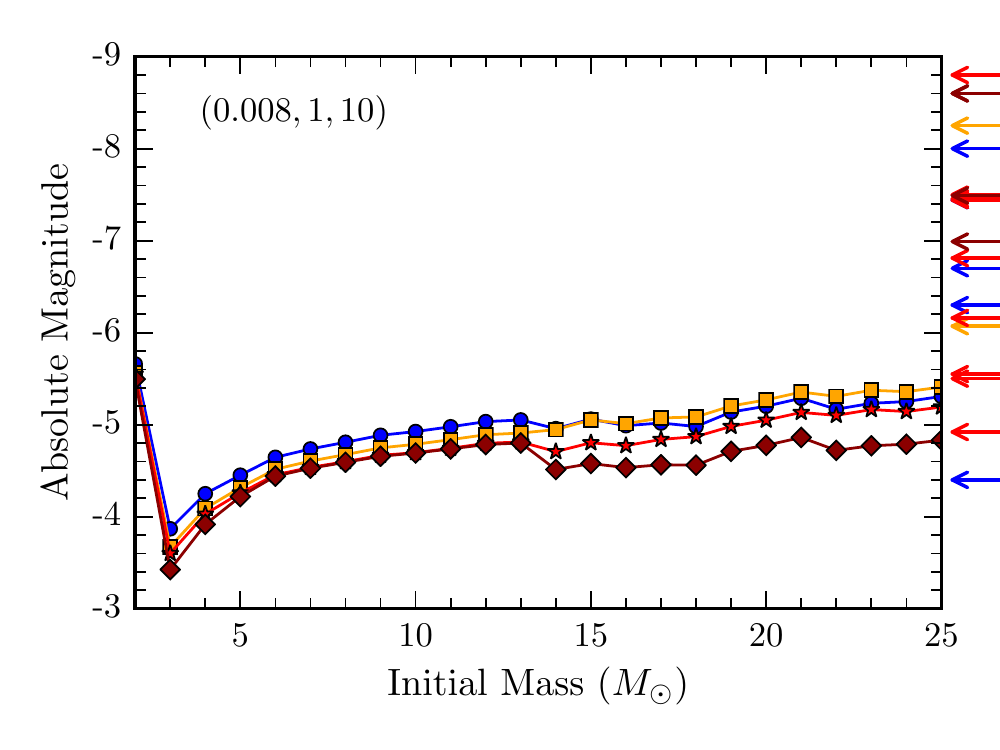}
	\includegraphics[width=\columnwidth]{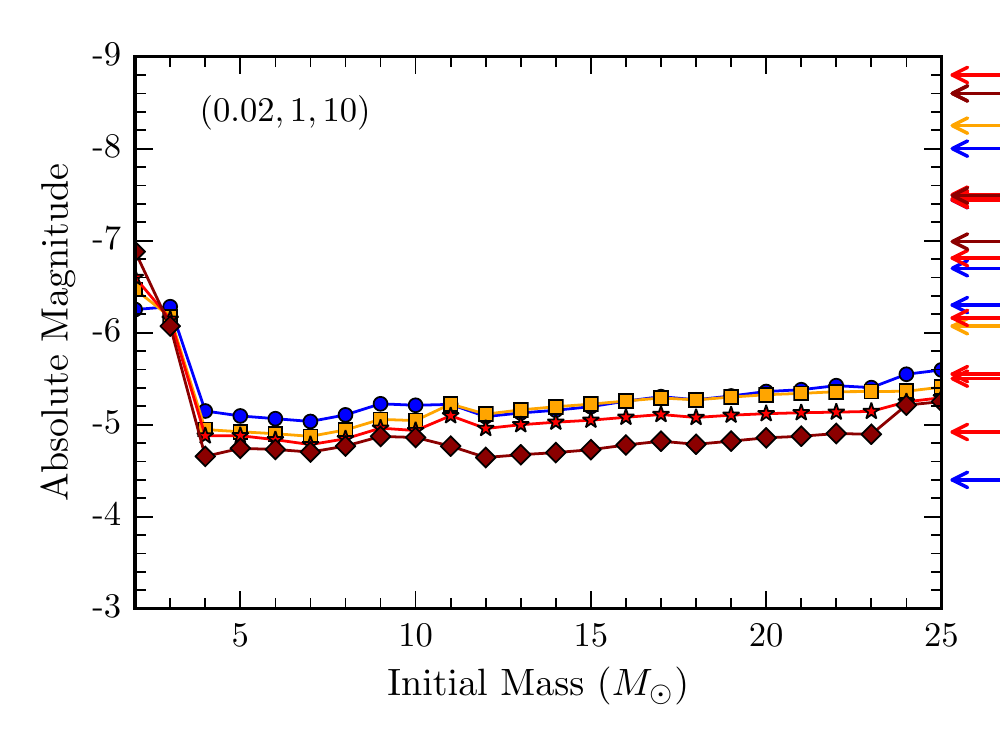}		
	\caption[]{Predicted supernova magnitudes of helium stars for models with $D=10$. The symbols retain their meaning from Fig.~\ref{fig:SNeMagsNoInflation}.}
	\label{fig:SNeMagsInflation}
\end{figure*}
Models with envelope clumping--that is, (0.008, 0, 10) and (0.008,
0.5, 10)--reproduce the late-type WN stars observations well. When a
higher mass-loss rate is used (i.e. $\beta=1$), we find poor agreement
with the late-type WN stars. We may argue that late-type WN stars in
the LMC have highly clumped envelopes with low mass-loss rates. In
contrast, the early-type WN stars in the LMC are likely to have
unclumped envelopes or a higher mass-loss rates.

The differences in evolution between mass-loss rate and metallicity
allow us to gain an independent estimate of the metallicity and
mass-loss history of these stars.

\subsubsection{LMC WC stars}

To compare our models with observed WC stars in the LMC, we have made
use of the results of \citet{2002A&A...392..653C}. As shown in
Fig.~\ref{fig:WRObsZ008}, the model set (0.008, 0.5, 10) is in good
agreement with the observed LMC WC star positions. As above, we see that only with reduced mass loss and the inclusion of clumping can we reproduce the observed WC stars locations on the HR diagram. Although we note the observed locations of the WO stars can be reproduced if we use a higher mass-loss rate. This indicates that, while WC evolution is the next evolutionary step for most WN stars, some WN stars experience more mass loss--either by being more massive, or by some other mechanism--to become a WO star.

The observations of these WC stars in the LMC were analysed using
\textsc{cmfgen} \citep[see][for details]{1998ApJ...496..407H}, and so
there may be a systematic difference in the derived parameters
compared to what might be found with the Potsdam code
(\textsc{powr}). In their analysis of the Galactic WC stars,
\citet{2012A&A...540A.144S} mentioned that, in a comparison between \textsc{cmfgen} and \textsc{powr}, there is very good agreement between the results from the codes with \textsc{powr} giving 0.1 dex higher luminosities.  

\subsection{The deaths of helium stars}

While charting the evolution of helium stars provides some degree of clarity as to their nature, the way in which they end their lives is, perhaps, equally as revealing.

We synthesise magnitudes data for our models using the \textsc{bpass} code \citep[for a complete description, see][]{2009MNRAS.400.1019E,2013MNRAS.436..774E}. In Figs.~\ref{fig:SNeMagsNoInflation} and \ref{fig:SNeMagsInflation}, we present the resulting progenitor magnitudes in the \textit{BVRI} bands. We immediately note a significantly higher visual magnitude for models without mass loss (i.e. $\beta=0$) than those with mass loss ($\beta=1$). 
This is expected as models with $\beta=0$ evolve to reach lower stellar temperatures due to inflation (see section \ref{ssec:evohigh}). Additionally, there is a clear correlation between initial mass and magnitude for models without mass loss above a threshold initial mass (approximately $4\msun$). However, for models with mass loss, no such trend is apparent. With mass loss, the high-mass helium stars evolve to obtain higher stellar temperatures than low-mass helium stars, as first identified by \citet{2012A&A...544L..11Y}. 
Though the high-mass helium stars are more luminous, the majority of their flux is output in the UV bands due to their higher surface temperatures. In contrast, the cooler temperatures of the evolved low-mass helium stars moves the peak of emission to longer wavelengths--closer to the visible part of the spectrum. In all cases shown, the lowest mass helium stars have the brightest progenitors regardless of the value of $\beta$. These are the helium giants, which are very cool and luminous; thus, like iPTF13bvn \citep{2013ApJ...775L...7C,2015MNRAS.446.2689E,2014AJ....148...68B}, they are easier to detect. 
The effect of envelope clumping on the magnitudes is very small: apart from the case of no mass loss, the final magnitudes of the models are comparable regardless of the value of $D$.

We now compare our results with the detection limits. The non-detection of Type Ib/c supernovae progenitors are discussed by \citet{2012A&A...544L..11Y} and \citet{2013MNRAS.436..774E}. Included on the right ordinates of Figs.~\ref{fig:SNeMagsNoInflation} and \ref{fig:SNeMagsInflation} are the detection limits from \citet{2013MNRAS.436..774E}. We posit the following about Type Ib/c supernovae and their progenitors.

Any population of helium or Wolf--Rayet stars must have a high probability of not being detected all the while explaining the formation of a luminous, low-mass progenitor. For models with $\beta=1$ and initial mass $>3\msun$, the progenitors lie outside nearly all of the detection limits. This is consistent with the lack of detection of Type Ib/c progenitors. The exception is the case of the helium giant, iPTF13bvn, which is consistent with a low-mass helium star ($<3\msun$). We can rule out the $\beta=0$ case for high-mass helium stars because otherwise they might have been observed. Helium giants with a low metallicity have comparatively fainter progenitor magnitudes than those with a higher metallicity. Therefore, helium giants might be observable only in metal-rich environments.

Upon comparing the non-detection limits discussed in \citet{2013MNRAS.436..774E} with the final end-points from our model grids, we find that nearly all the models would be undetected in the pre-explosion imagery. The only case that should have been detected is SN2002ap: a Type Ic supernova with a low-mass ejecta. This result is not too disagreeable, as the explosion may have destroyed an unknown amount of extra extinction in the system. Additionally, any extra mass loss would further reduce the luminosity of the progenitors while increasing the surface temperature, making a pre-explosion detection more difficult \citep[][]{2015arXiv150500270T}. As is the case for our models, no helium-giant-type model is found to have been stripped of its envelope. Therefore, a star that has been stripped of its outer envelope will be much fainter than what our models predict.

The prediction that the most luminous progenitors are, in fact, those of the lowest mass was first noted by \citet{2012A&A...544L..11Y}. It is, therefore, expected that helium giant Type Ib/c progenitors will be observable, while progenitors from WR stars will remain difficult to detect. However, we propose that these WR-star progenitors are not as faint as those suggested by \citet{2012A&A...544L..11Y} and \citet{2013A&A...558A.131G}. We find brighter progenitors as our stars cool towards the end of their evolution, and so more radiation is emitted in the optical bands than these hotter models. Thus, we predict the lowest mass progenitors will not only have ejecta masses similar to those of iPTF13bvn, but will also match the revised observed magnitudes from \citet{2015MNRAS.446.2689E}.

Furthermore, we note that there is little dependence between the final magnitudes and the initial (or final) mass of the helium star. Therefore, with the detection of a Type Ib/c supernova, it becomes possible to classify the nature of the progenitor: either helium giant or WR. However, an exact determination of the initial helium star mass or the initial stellar mass is impossible.

We may use the models to predict the evolutionary end-point of stars as a function of their initial mass. Fig.~\ref{fig:compoIMF} examines the effects of metallicity and mass loss on helium star models. Models with an initial mass $\le 2\,\msun$ terminate with a highly degenerate CO core. Above $2\,\msun$, the binding energy of the envelope is sufficient to result in a core-collapse supernova \citep[][]{2004MNRAS.353...87E}. The composition of the ejecta depends on the initial mass of the model. Models with an initial mass $\ga10\msun$ explode with a small helium presence in their ejecta ($\sim 0.2\,\mathrm{M}_{\odot,\mathrm{He}}$). From this, we may infer that high-mass helium stars are potential candidates for helium-poor Type Ic supernovae \citep[see][]{2015arXiv151008049L}. Models with a low initial mass ($\la10\msun$) do not evolve with significant mass loss, and therefore, are unable to remove their helium envelope before death. As such, their ejecta composition is helium rich; a potential candidate for Type Ib supernovae. Though these results appear conclusive, the determination of supernova type from a stellar model is an inherently complex adventure \citep[see][]{2011MNRAS.414.2985D,2014ApJ...792L..11P,2012MNRAS.424.2139D,2015PASA...32...15Y,2013ApJ...773L...7F}.

\section{Discussion \& Conclusions}
\label{sec:discussion}

By comparing different sets of stellar models, we have determined two important factors affecting the evolution of WR stars. They are the metallicity and mass-loss rate used. By varying the input physics, we have created models that reach all the locations of observed WR stars in the HR diagram; however, we must ask ourselves how likely these different model sets are. Without a full, detailed synthetic population that includes the evolution prior to the HeZAMS, it is difficult to draw a definitive conclusion. None the less, we may deduce some important facts from our models.

We note that our models, in terms of evolutionary pathways on the HR diagram (Figs.~\ref{fig:WRObs} and \ref{fig:WRObsZ008}) and predicted lifetimes (Fig.~\ref{fig:lifetimes_ML}) for the stars, compare well qualitatively with other similar models. For example, those by \citet{1986A&A...167...61H,1998NewA....3..443V,1999A&A...350..148W,2002MNRAS.331.1027D,2008MNRAS.384.1109E,2012A&A...540A.144S,2013A&A...558A.131G,2015A&A...580A..20S}. The only significant difference between models is with the high-mass helium stars, which move on to the WR hook (or loop) during the WO phase following core-helium burning. Different evolutionary codes appear to react differently after reaching the hottest point of WO evolution. Our models, for instance, evolve to be significantly cooler than other models. Only, therefore, with observational data of a WO star as a SN progenitor will the correct evolution be identified. However, this WR hook is a common feature in all but one of the helium star models, so there is consensus that this is, indeed, an evolutionary feature.

One immediate fact we can discern from our results is that the magnitudes of current mass-loss rates employed in WR star models are roughly correct. In all cases where we increase the mass-loss rate, the $\beta=2$ models, we find that none reach positions in the HR diagram close to where WR stars are observed. We are also able to conclude that having dramatically lower mass-loss rates, e.g. our $\beta=0$ models, are required for some observed WR stars, but not all. 
Although, we find evidence that suggests the mass-loss rates used for LMC WR stars should be reduced.

We now consider the evolutionary natures of the WR stars we observe. The Conti scenario describes the expected evolution: WN stars go from early to late types and are succeeded by a WC phase, while the most massive stars end with a WO phase. Current observational evidence casts doubt on this scenario. The observed luminosity and temperature ranges of WC and WO stars are inconsistent with evolutionary models. The hydrogen-free, late-WN stars are problematic as most stellar models at their locations still contain hydrogen. However, as we have noted in section~\ref{ssec:GalacticWN}, the only way to reproduce the late-WN stars with evolutionary models is either by reducing the mass-loss rate so that a WR star has a more massive helium envelope, or by including clumping in the inflated envelope.

A proposed explanation for these hydrogen-free WNL stars is inflation of the envelope. A WR star spends most of its lifetime burning helium (see Fig.~\ref{fig:lifetimes_ML}) and is, therefore, expected to be WNE. \citet{2012A&A...538A..40G} created models suggesting the envelopes of these stars can be inflated; the radii of stellar evolution models are too small by a significant factor. This is in agreement with the detailed atmospheric models calculated by \citet{2013A&A...558A.131G} for the Geneva stellar evolution models; hydrogen-free WN stars should be much hotter than we observe them to be. Inflation may solve this problem; however, such inflation is likely dependent on the mass-loss history of the star. A star that evolves with little mass loss will have a comparatively larger helium envelope, and therefore, a larger radius. Conversely, a greater amount of mass loss will result in a smaller helium envelope, and thusly, a smaller star. Although, as stars in the LMC indicate towards lower metallicities, we may enhance the inflation by allowing the envelope to be clumped.

There also exists the possibility of additional physical effects that increase the effective radii of the stellar models not considered by the evolutionary code. Effects such as rapid rotation that decreases the surface gravity; however, most models suggest WR stars rapidly lose angular momentum. Otherwise, inflation is the most likely candidate and, while it does occur in our models, the effect is much weaker than is required at the correct metallicity. However, as above, we may augment inflation with an inhomogeneous envelope, thereby increasing the radii of our WR models.

In light of our work, we can draw some firm conclusions about certain aspects of WR star evolution and speculate about others. We shall now discuss our conclusive findings.

First, WO stars are what we have always considered to be WC stars in stellar models. They are the progeny of the most massive WN stars ($M_\mathrm{He,i}>8\msun$) that have suffered significant mass loss and are the hottest WR stars. Due to their significant mass loss, WO stars are likely to explode as Type Ic supernovae at any metallicity. However, these massive stars are also likely to form black holes at core collapse, so it is unknown as to whether they produce visible supernovae \citep[e.g.,][]{2015PASA...32...16S}. 

Second, WC stars evolve from less massive stars ($M_\mathrm{He,i}<8\msun$). The evolution of these stars could be described either as an inflationary effect occurring towards the end of their lives or as them becoming helium giants. The WC stars experience increased mass loss as they evolve; a consequence of a decrease in surface gravity allowing material on the surface to be removed more efficiently. The expansion of the envelope, whether through inflation or a helium-giant phase, is metallicity dependent. For low metallicities, the stars would retain a small fraction of hydrogen in their envelopes precluding any expansion. This is in agreement with the lack of observed WC stars at low metallicity.

We note the WC stars are unlikely to be the evolutionary end-points of the typical WN stars observed. They are more likely to arise from lower mass objects that we have yet to find; the recently identified and very faint WN3/O3 stars discovered by \citet{2015IAUS..307...64M} are also possible candidates. 
Though the WN3/O3 stars may indeed be the progenitors of some of the WC stars, the WC progenitors are about 0.3 dex fainter, hotter, and less luminous than those typically observed for WR stars during their core-helium burning lifetimes. This means that during most of their existence they are unlikely to have the stellar wind emission lines of their more luminous relatives, making them difficult to identify. Furthermore, they are most likely to exist only in binary systems in orbit around cooler and more luminous O-stars, making them almost impossible to detect. We suggest that the only way they may be observed is if their high temperatures were to give rise to highly excited nebular emission lines, such as He\,{\sc ii}. This could explain such He\,{\sc ii} nebulae where no WR stars have been observed \citep[e.g.][]{1991ApJ...373..458G,2011A&A...526A.128K}.

The identification of WC and WO stars coming from different mass ranges is likely as the post-helium-burning lifetime of massive helium stars compared with the evolution of the lower mass stars is similar, so the relative numbers may be explained by a similar evolutionary time scale. Recent work by \citet{2015MNRAS.447..598S} on luminous blue variable (LBV) stars provides additional evidence for WC stars being less massive than WN stars.

Further evidence of this can be inferred from the metallicity evolution. We see from the models the luminosity of stars that become WC stars increases at $Z=0.008$ compared with $Z=0.020$. This is also apparent in the observations. Also, the temperature distribution becomes much tighter at low metallicity with no cool WC stars possible from our tracks; at the higher metallicity, cool WC stars are possible. This fact has interesting implications for SN progenitors. Whereas in the Galaxy WR stars are most likely to be either WC or WO stars at the end of their evolution, in the LMC they can be WN, WC or WO depending on their initial mass.

Our findings on WN stars prove much more speculative, and we provide an alternative to the enhanced inflation suggested by others \citep[see][]{2012A&A...538A..40G,2006A&A...450..219P,1999PASJ...51..417I}. Although it is likely inflation still occurs (as evidenced by models with $\beta=0$), it is possible that some hydrogen-free WN stars, especially late-type WN stars, are pre-helium main-sequence stars. While the evolutionary time scales will be comparatively short, they are similar to the post-helium-burning time scales for the eventual stars. If this is the case, then only the hotter (earlier) WN3 and WN2 stars are, in fact, core-helium-burning stars. We might expect, due to their longer lifetimes, a larger population of such stars; however, these PHeMS stars are not observed in the Galaxy. A recent study of the LMC by \citet{2015IAUS..307...64M} revealed a new class of WR stars, WN3/O3, that are hot and at locations similar to those expected of helium-burning WN stars. Because their absolute visual magnitudes are extremely low, they are difficult to detect; none the less, their discovery corresponds to a 6 per cent increase in the number of known WR stars in LMC. 

From these suggestions, we note some important consequences. For WC stars, the models still have nitrogen on their surfaces. In the case of very late WC9 stars, the removal of surface nitrogen may be facilitated by extra mixing (for example, from rotation), removing the nitrogen through nuclear processing rather than mass loss.
If the stellar envelope is clumped, it can only occur for WN stars. The only way to reproduce the locations of the late-WN stars on the HR diagram with standard-evolution models is to remove all mass loss, as even slight mass loss will drive a model to higher stellar temperatures. WC and WO stars do not require enhanced inflation to reproduce their HR diagram positions. The evolutionary models predict the locations of WO stars without a problem: a hot phase consequential to high-mass WN stars.
The predictions for WC stars are correctly realised for the cooler temperatures of low-mass WC stars. In truth, it is possible both points are at play and only through future observations are they likely to be clarified.

Finally, for the transition WN/WC stars in the Galaxy, two are in the expected location for such transition objects--assuming clumping is included, and in one case, the mass-loss rates are reduced. The most luminous transition object is more massive than the models we present, but could well evolve in the future towards a WO star. A population synthesis will need to consider the entire evolution of these stars and how they might be affected by rotation or binary interactions.

In this work, we have created models that evolve to reach portions of the HR diagram where the observed WR stars are located. To reproduce the observed locations, our models require the inclusion of clumping or a mass-loss rate of zero. However, whether these models are representative of the stellar population remains to be seen. The stars are likely to arise from a confluence of origins: non-rotating and rapidly rotating single stars, as well as interacting binaries. The lower mass systems are more favourable of binary evolution because stellar winds at lower luminosities are likely to be weaker, unable to strip the hydrogen envelopes. By considering the evolution of helium stars without the complication of their earlier evolution, we have gained new insights into their evolution. In addition, we have created a new set of stellar models--freely available to the astronomical community--that, to our knowledge, have not previously existed.

\section{Acknowledgements}

LASM and JJE thank Robert Izzard for automating the process of obtaining the Potsdam WR models, and thank Mason Ng and Georgie Taylor for their work on the locations of massive stars in the Galaxy. JJE acknowledges support from the University of Auckland. The authors wish to acknowledge the contribution of the NeSI high-performance computing facilities and the staff at the Centre for eResearch at the University of Auckland. New Zealand's national facilities are provided by the New Zealand eScience Infrastructure (NeSI) and funded jointly by NeSI's collaborator institutions and through the Ministry of Business, Innovation and Employment Infrastructure programme\footnote{\texttt{http://www.nesi.org.nz}}. We also thank the anonymous referee for their constructive comments that have improved this paper.

\bibliographystyle{mn2e}
\bibliography{article_helium}

\clearpage
\setcounter{figure}{6}

\begin{figure*}
	\centering
	\includegraphics[width=\columnwidth]{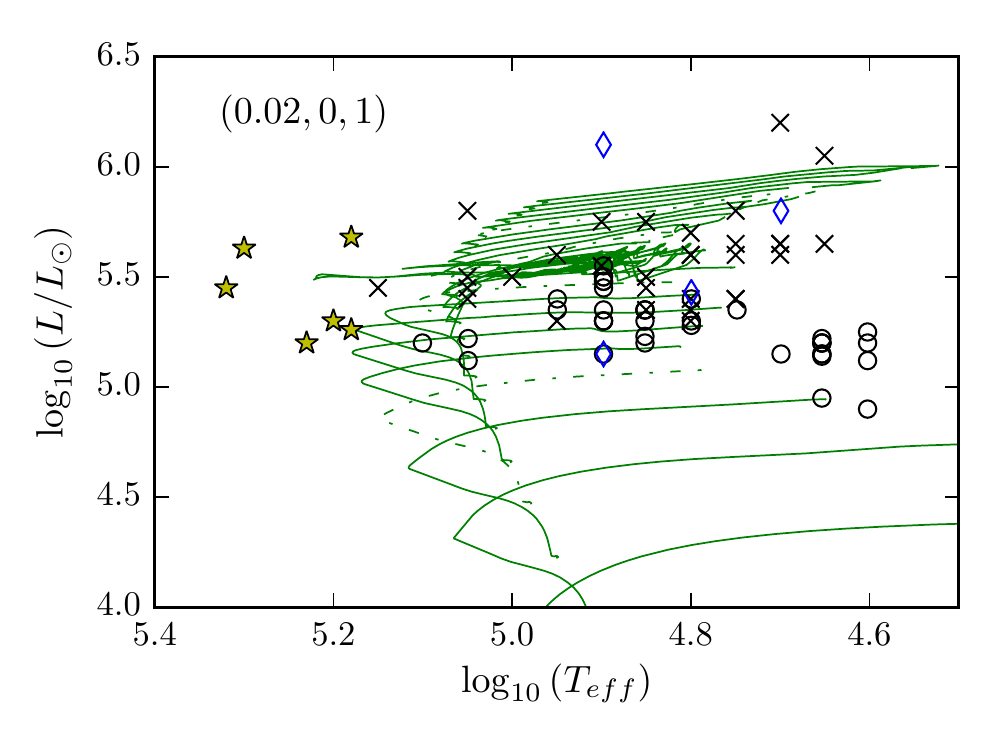}	
	\includegraphics[width=\columnwidth]{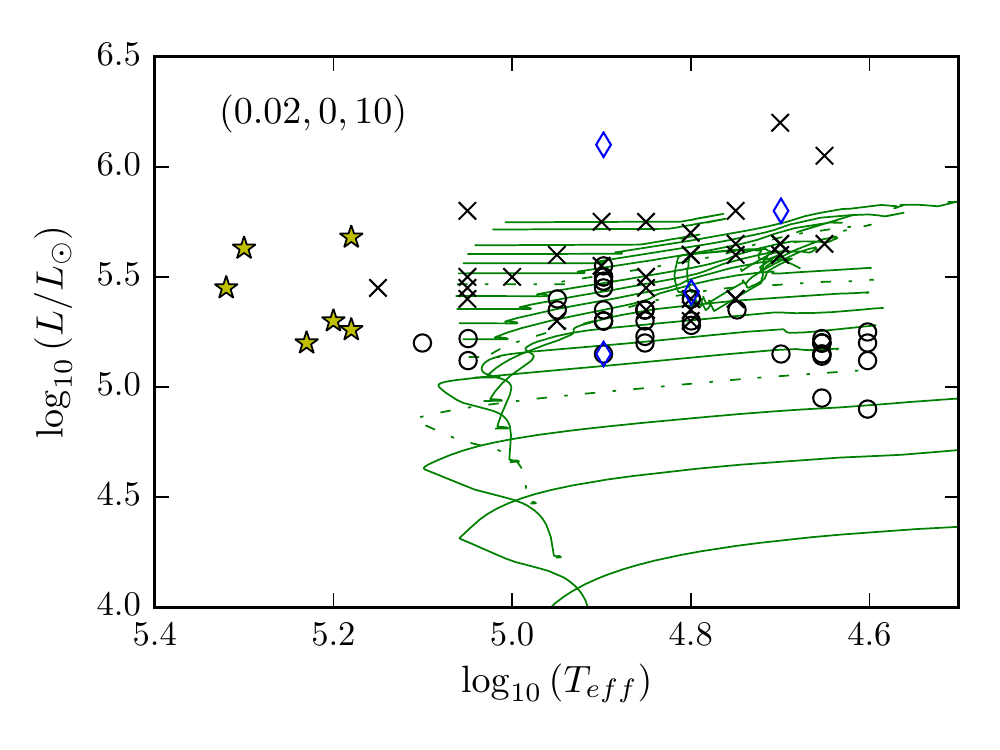}\\
	\includegraphics[width=\columnwidth]{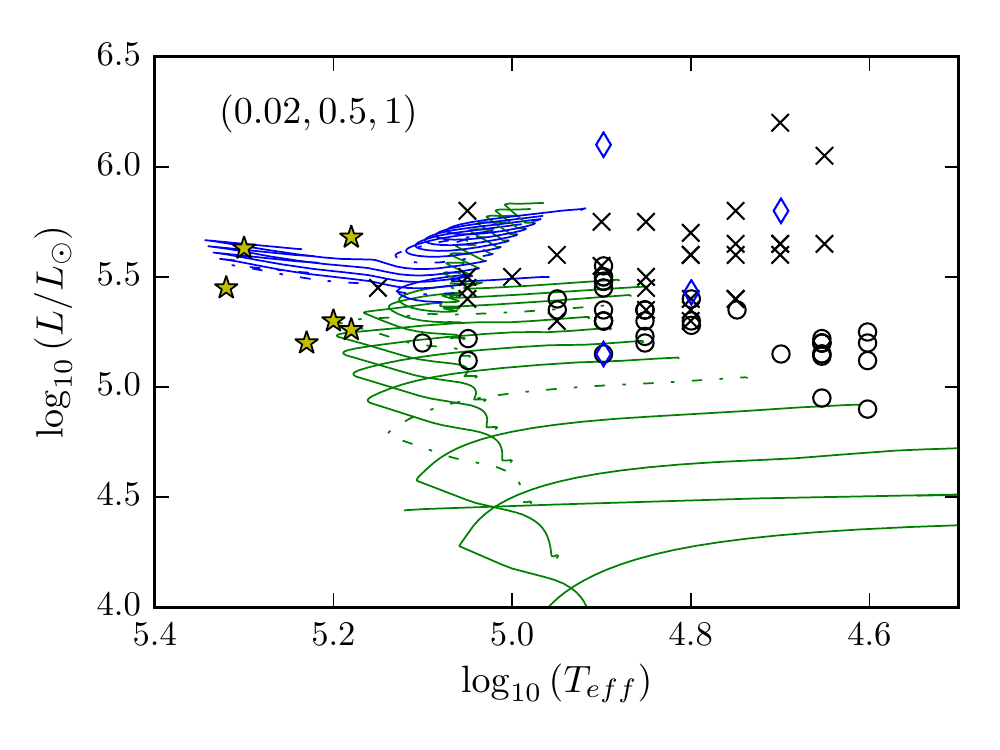}			
	\includegraphics[width=\columnwidth]{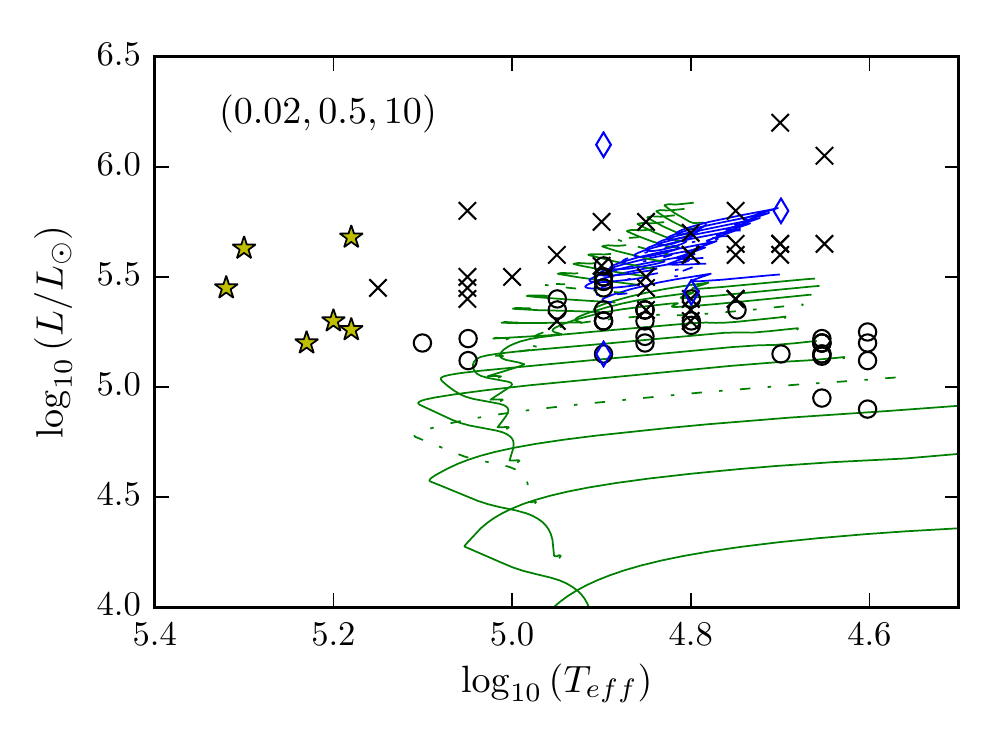}\\
	\includegraphics[width=\columnwidth]{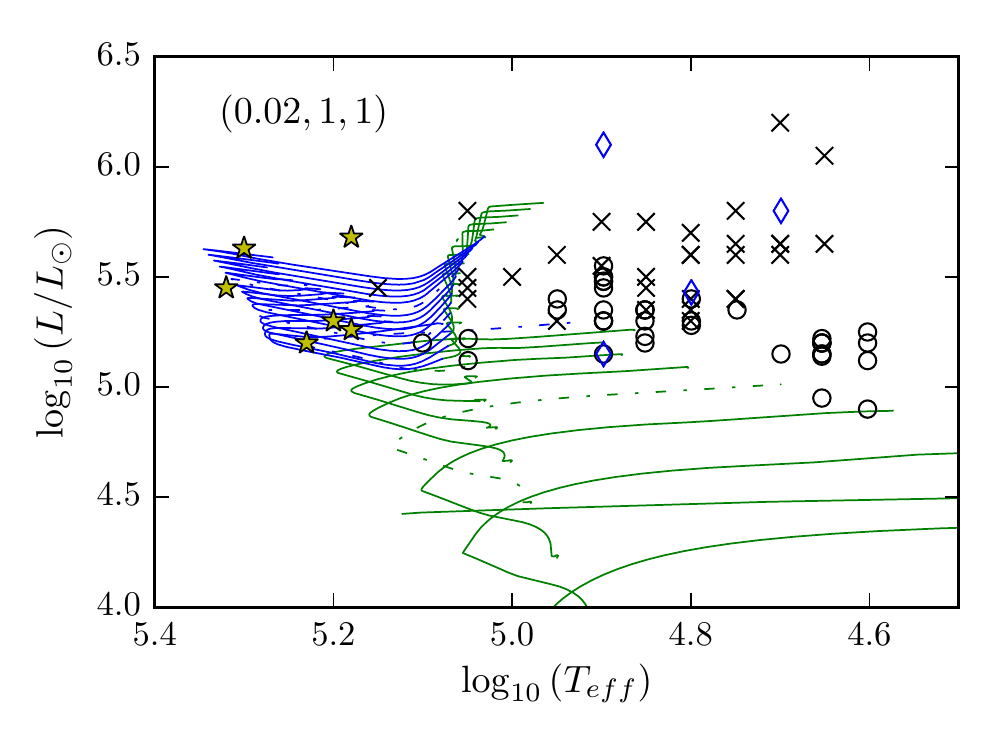}
	\includegraphics[width=\columnwidth]{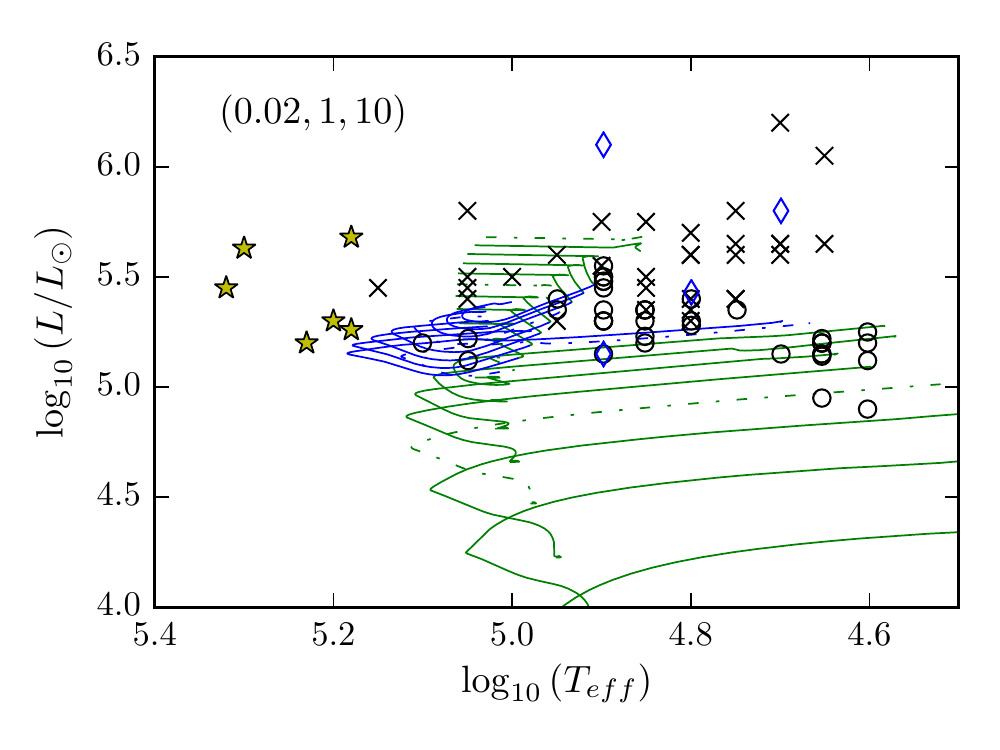}			
		
	\caption[]{HR diagram of our evolved models. For clarity, models with initial masses of $5,10,15 \ \mathrm{and} \  20\msun$ are plotted with broken-dotted lines. Observed WR star locations are marked as follows: WN, saltires; WC, circles; WO, yellow stars; and WN/WC transition objects, blue diamonds. Black represents Galactic observations, while red represents observations in the LMC. All observed stars are hydrogen-free. For details and references, see section~\ref{ssec:obswrstars}.The phase of WR mass loss is indicated on the tracks as follows: WN, green; and WC, blue. Metallicity, mass-loss rate, and clumping factor are noted on the plots.}
	\label{fig:WRObs}
\end{figure*}

%
%
%

\begin{figure*}
	\centering
	\includegraphics[width=\columnwidth]{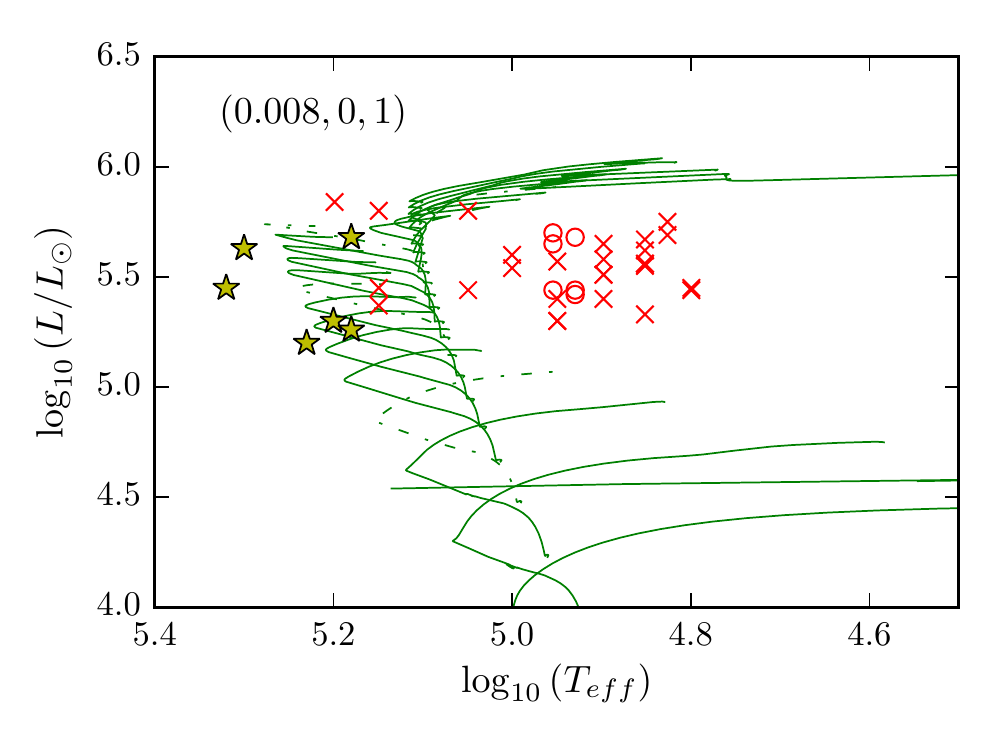}	
	\includegraphics[width=\columnwidth]{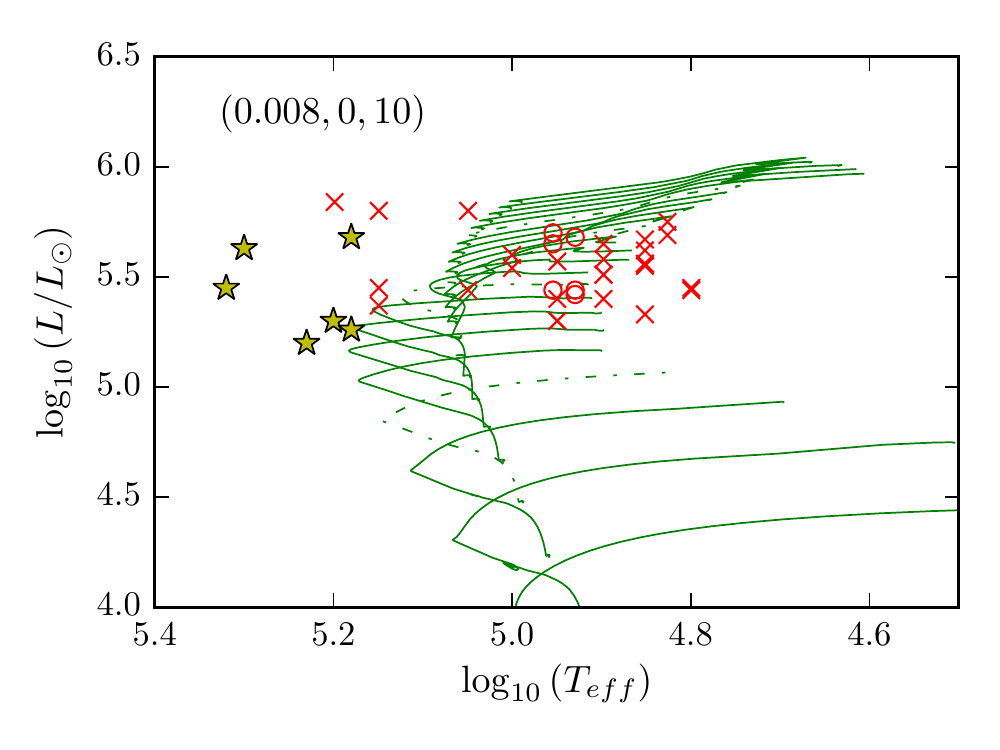}\\
	\includegraphics[width=\columnwidth]{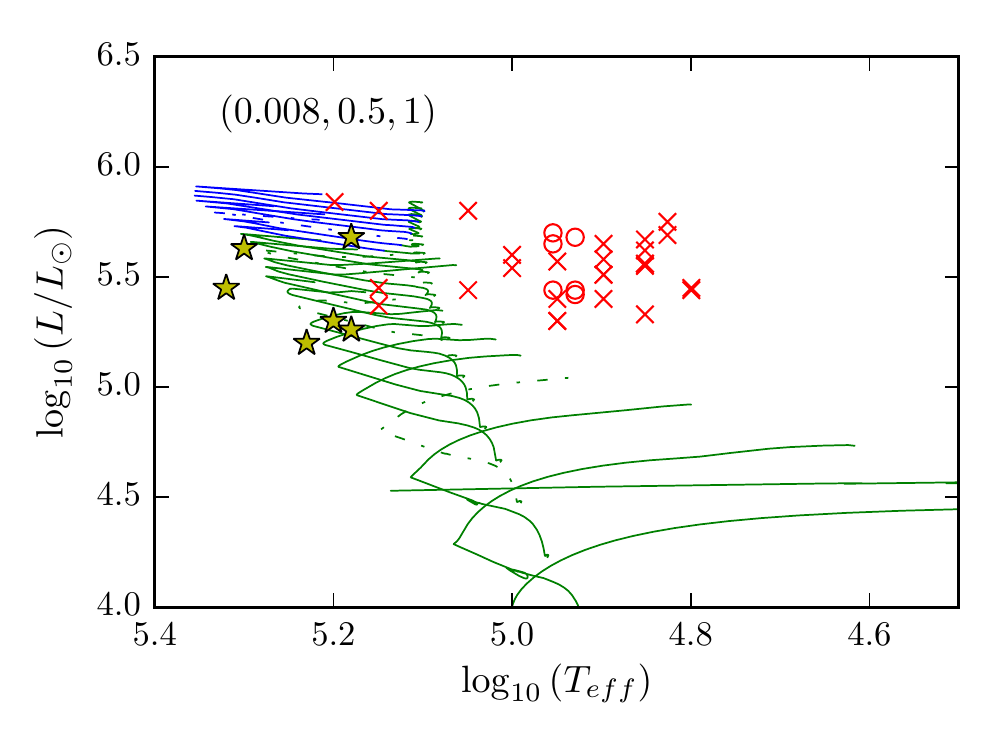}			
	\includegraphics[width=\columnwidth]{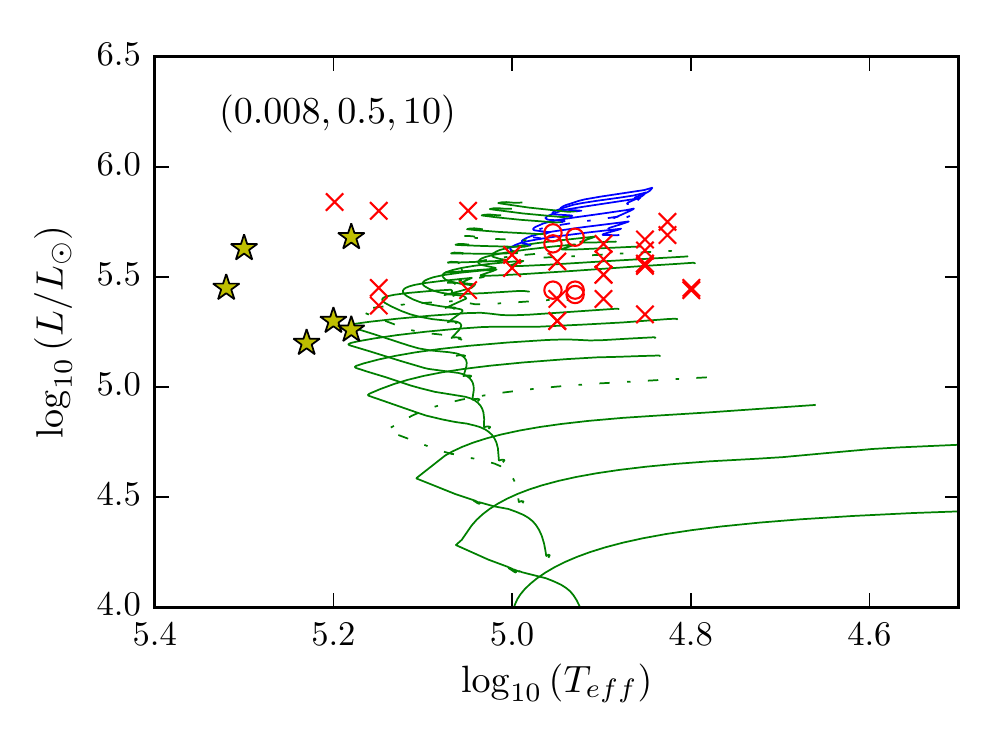}\\	
	\includegraphics[width=\columnwidth]{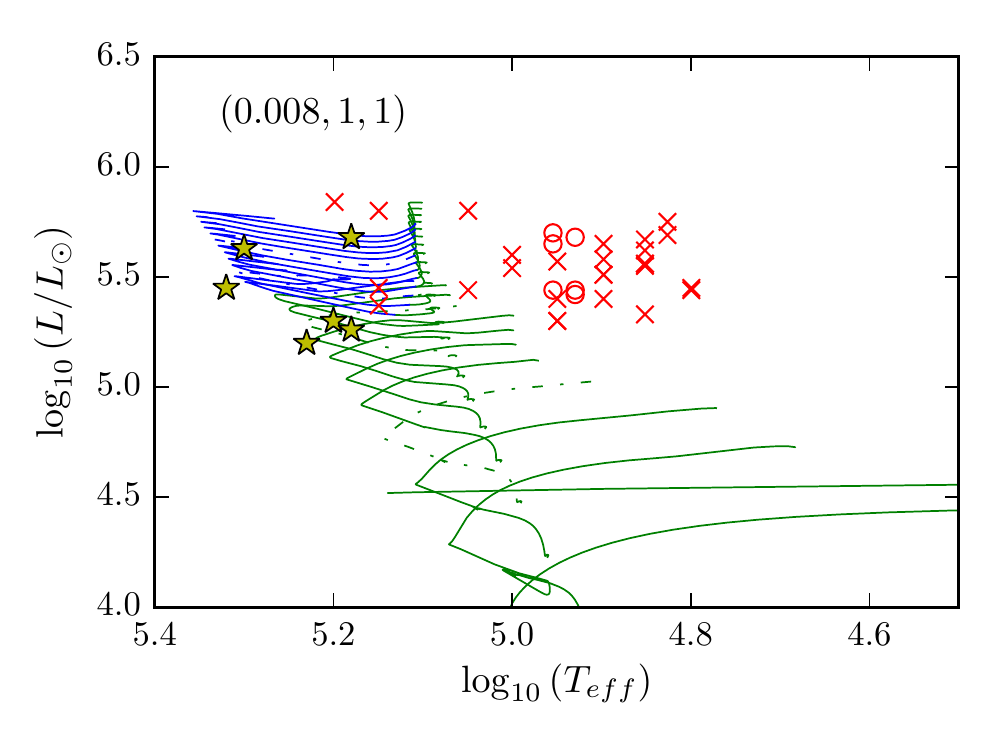}
	\includegraphics[width=\columnwidth]{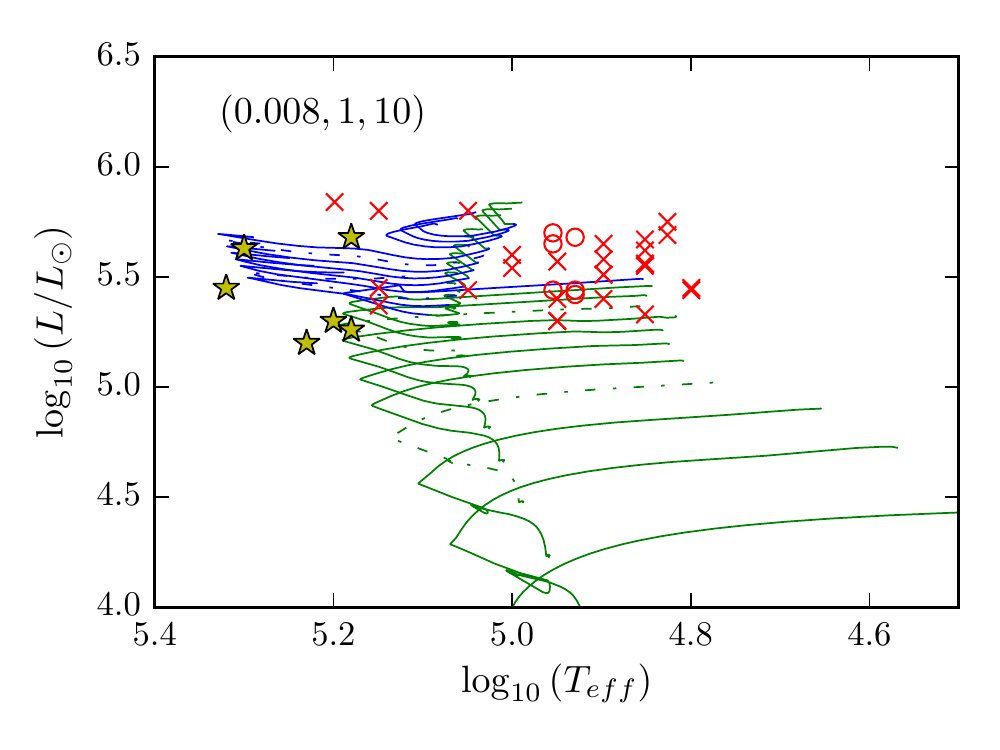}			
	\caption[]{Same as Fig.~\ref{fig:WRObs}, but for $Z=0.008$ models.}
	\label{fig:WRObsZ008}
\end{figure*}
%
%
%
\setcounter{figure}{12}

\begin{figure*}
	\centering
	\includegraphics[width=\columnwidth]{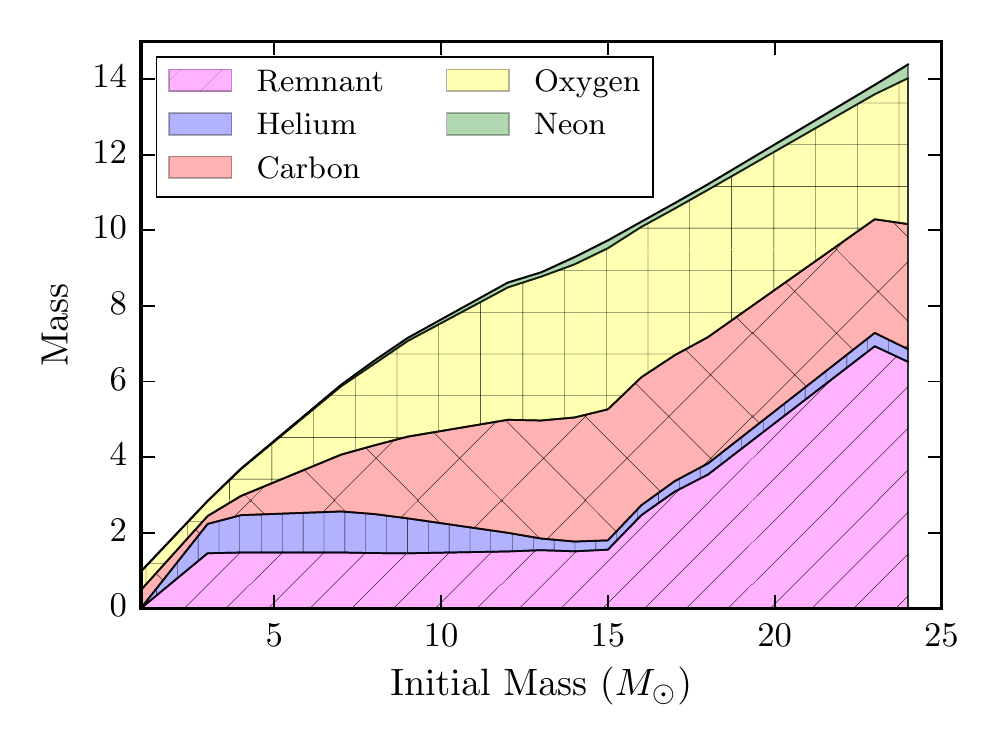}
	\includegraphics[width=\columnwidth]{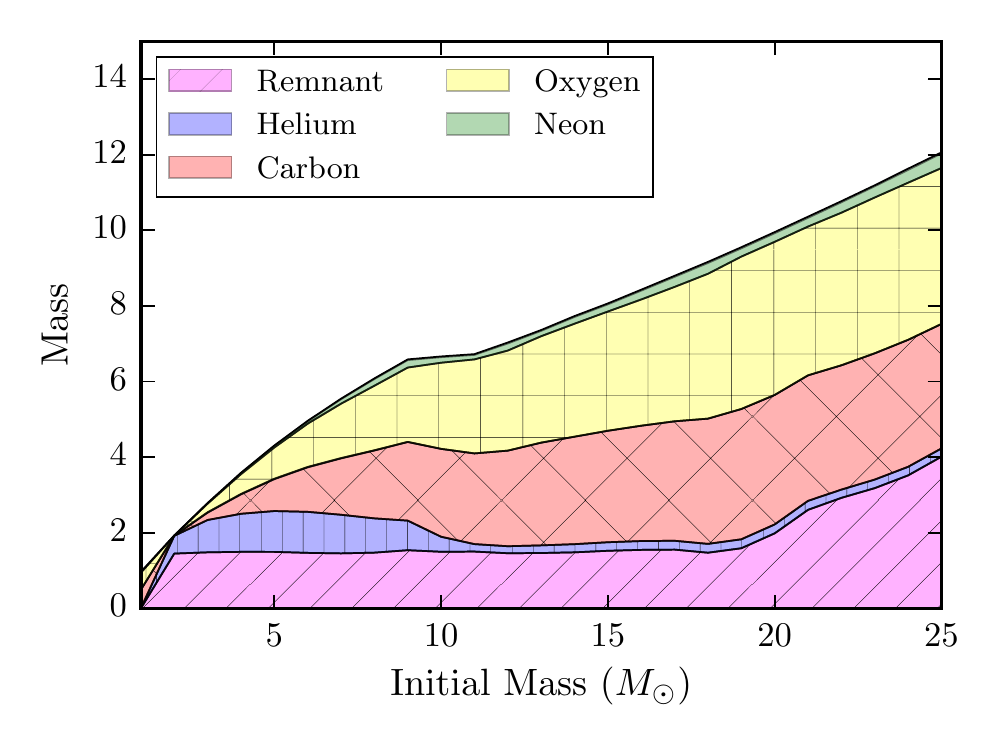}\\
	\includegraphics[width=\columnwidth]{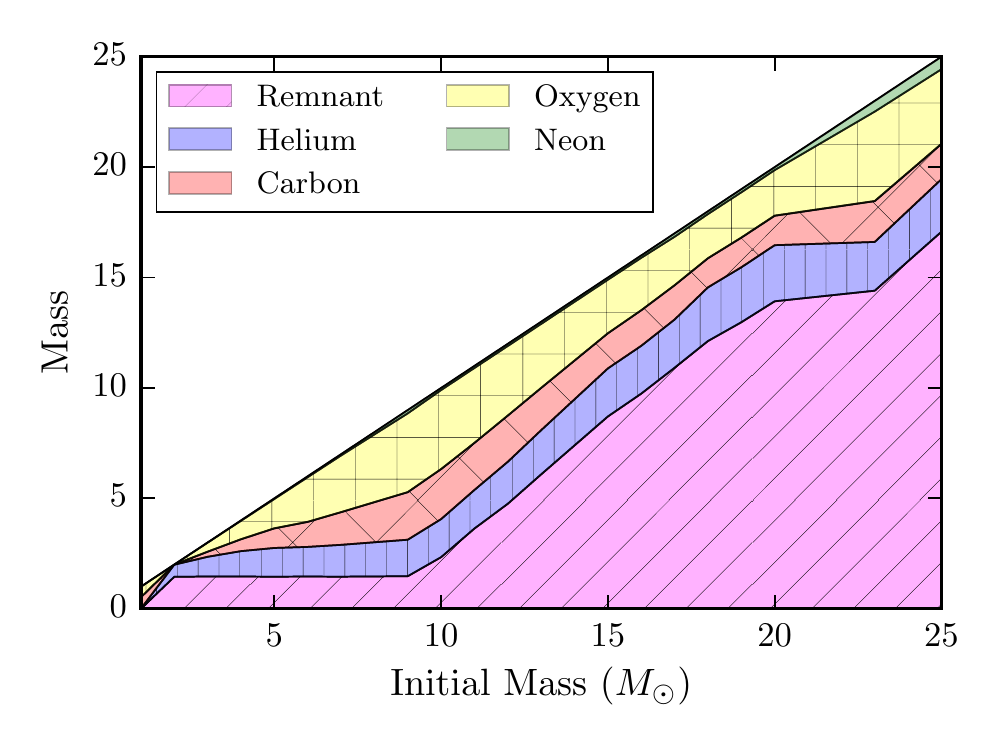}
	\includegraphics[width=\columnwidth]{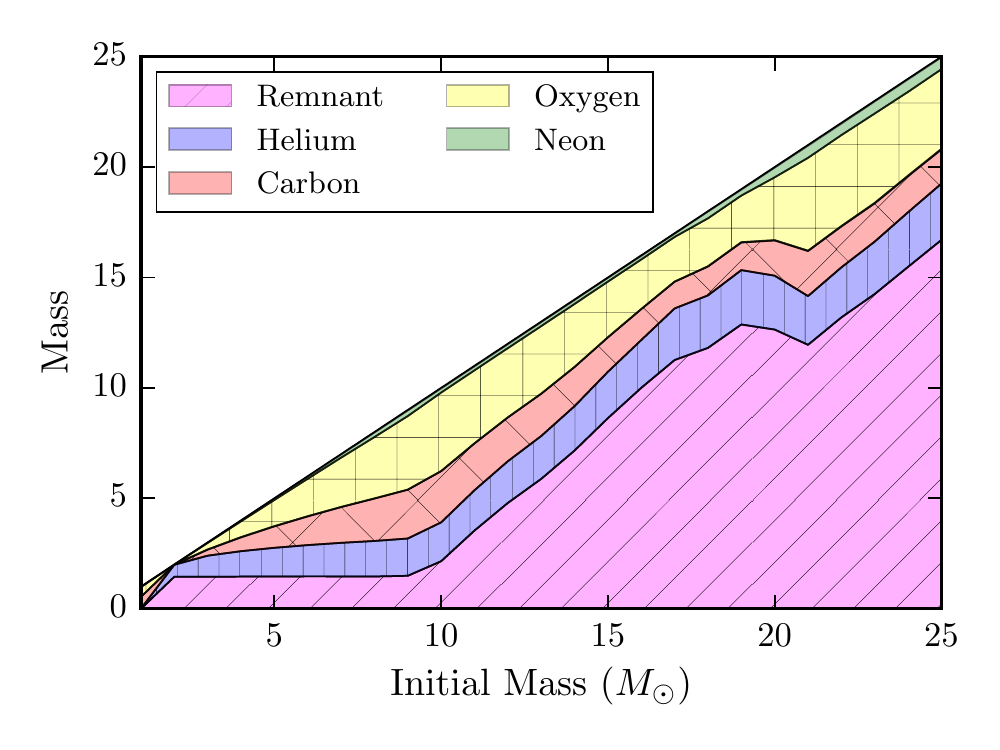}	
	\caption[]{The final composition of helium star models at $Z=0.008$ (\textit{left column}) and $Z=0.02$ (\textit{right column}) as a function of initial mass. The top row displays models with $\beta=1$, while the bottom row shows models with $\beta=0$. The ordinate gives the mass of the coloured quantity for a model of initial mass given by the abscissa.}
	\label{fig:compoIMF}
\end{figure*}

\end{document}